%% file: main.tex
\documentclass[conference]{IEEEtran}
\IEEEoverridecommandlockouts
\usepackage[T1]{fontenc}
\usepackage{float}
\usepackage{tabularx}
\usepackage{algorithm}
\usepackage{algpseudocode}
\usepackage{xcolor}
\usepackage{amsmath}
\usepackage{tikz}
\usepackage{graphicx}
\usepackage{caption}
\usepackage{url}
\usepackage{subcaption}
\usepackage{booktabs}
\usetikzlibrary{decorations.pathreplacing, calc}
\usepackage{ipdps26repro}

\usepackage{booktabs}
\usepackage{tagging}
\usepackage{textcomp}

\def\BibTeX{{\rm B\kern-.05em{\sc i\kern-.025em b}\kern-.08em
    T\kern-.1667em\lower.7ex\hbox{E}\kern-.125emX}}

\usepackage{amsmath,amssymb,amsfonts}

\usepackage{cite}

\usepackage{graphicx}
\usepackage{tikz}
\usetikzlibrary{decorations.pathreplacing, calc}

\usepackage{booktabs}
\usepackage{tabularx}

\usepackage{algorithm}
\usepackage{algpseudocode}

\usepackage{xcolor}

\usepackage[caption=false,font=footnotesize]{subfig}

\usepackage{float}

\def\BibTeX{{\rm B\kern-.05em{\sc i\kern-.025em b}\kern-.08em
    T\kern-.1667em\lower.7ex\hbox{E}\kern-.125emX}}

\usetag{explanation}
\usetag{example}

\begin{document}

\title{From Skew to Symmetry: Node-Interconnect Multi-Path Balancing with Execution-time Planning for Modern GPU Clusters\thanks{This research is supported in part by NSF grants \#2311830, \#2312927,
\#2323116, \#2415201, \#2504944, and XRAC grant \#NCR-130002.}
}

\author{
\IEEEauthorblockN{
Jinghan Yao,
Kaushik Kandadi,
Bharath Ramesh,
Hari Subramoni,
Dhabaleswar K. Panda
}
\IEEEauthorblockA{
Department of Computer Science and Engineering\\
The Ohio State University\\
Columbus, OH, USA\\
\texttt{\{yao.877, kandadisuresh.1, ramesh.113, subramoni.1\}@osu.edu}\\
\texttt{panda@cse.ohio-state.edu}
}
}

\maketitle

\input{content/0-abstract}


\input{content/1-introduction}
\input{content/2-background_motivation}

\input{content/3-Overview_of_imbalanced}
\input{content/4-design}

\input{content/5-Evaluation}

\input{content/6-Related_works}
\input{content/7-Limitations_FutureWork}

\input{content/8-Conclusion}


\bibliographystyle{IEEEtran}
\bibliography{main}

\end{document}

%% file: content/0-abstract.tex
\begin{abstract}
Modern GPU-based high-performance computing (HPC) clusters offer unprecedented communication bandwidth through heterogeneous intra-node (e.g., NVLink, NVSwitch, AMD xGMI) and inter-node (e.g., multi-rail InfiniBand, RoCE NICs) interconnects. However, despite ample aggregate bandwidth, many important communication patterns in real-world applications routinely fail to utilize these hardware capabilities fully. Traffic skew frequently leads to scenarios where a small number of links become oversaturated while others remain largely idle, causing congestion, latency spikes, and severe scalability bottlenecks. Current communication frameworks (e.g., NCCL, MPI with UCX) predominantly rely on static fastest-path routing or static hashing-based multi-rail striping, leaving significant bandwidth unused whenever real-time load deviates from the expected distribution.

To overcome these limitations, we propose \textbf{NIMBLE} (\textit{Node-Interconnect Multi-path BaLancing with Execution-time orchestration}), a runtime communication orchestration system that adaptively redistributes network traffic to balance link utilization across all available intra-node and inter-node paths. NIMBLE solves a capacity-normalized minimum-congestion optimization via a fast multiplicative-weights scheme, and leverages CUDA-aware GPU kernel-based RDMA pipelining to efficiently route traffic through intermediate GPUs and rail-matched NICs. This endpoint-driven orchestration integrates transparently with existing communication libraries without application changes while preserving ordering, determinism, and low overhead.

On H100-SXM4 nodes with fully connected NVLINK and four NDR400 rails, NIMBLE achieves up to $2.3\times$ higher intra-node bandwidth and $3.8\times$ higher inter-node throughput compared to single-path baselines. It outperforms NCCL and MPI by up to $5.2\times$ on skewed All-to-Allv workloads, $1.35\times$ on the end-to-end LLM MoE blocks, while matching baseline performance under balanced traffic.
\end{abstract}

%% file: content/1-introduction.tex
\section{Introduction}
High-performance computing (HPC) and artificial intelligence (AI) workloads increasingly depend on large-scale clusters of GPUs interconnected by sophisticated, heterogeneous fabrics. Modern computing nodes typically integrate high-bandwidth intra-node communication paths (e.g., NVLink, NVSwitch, xGMI) \cite{nvidia-nvlink,nvidia-nvlink4-nvswitch,amd-mi300x-datasheet} and multiple inter-node network interface cards (NICs) employing InfiniBand or RoCE protocols \cite{nvidia-ndr-overview,cisco-rocev2}, collectively providing terabytes per second of aggregate bandwidth per node. Ideally, applications should exploit this rich connectivity to achieve high communication throughput and low latency, ensuring scalability and efficient resource usage.

\begin{figure}[t]
    \centering
    \includegraphics[width=0.47\textwidth]{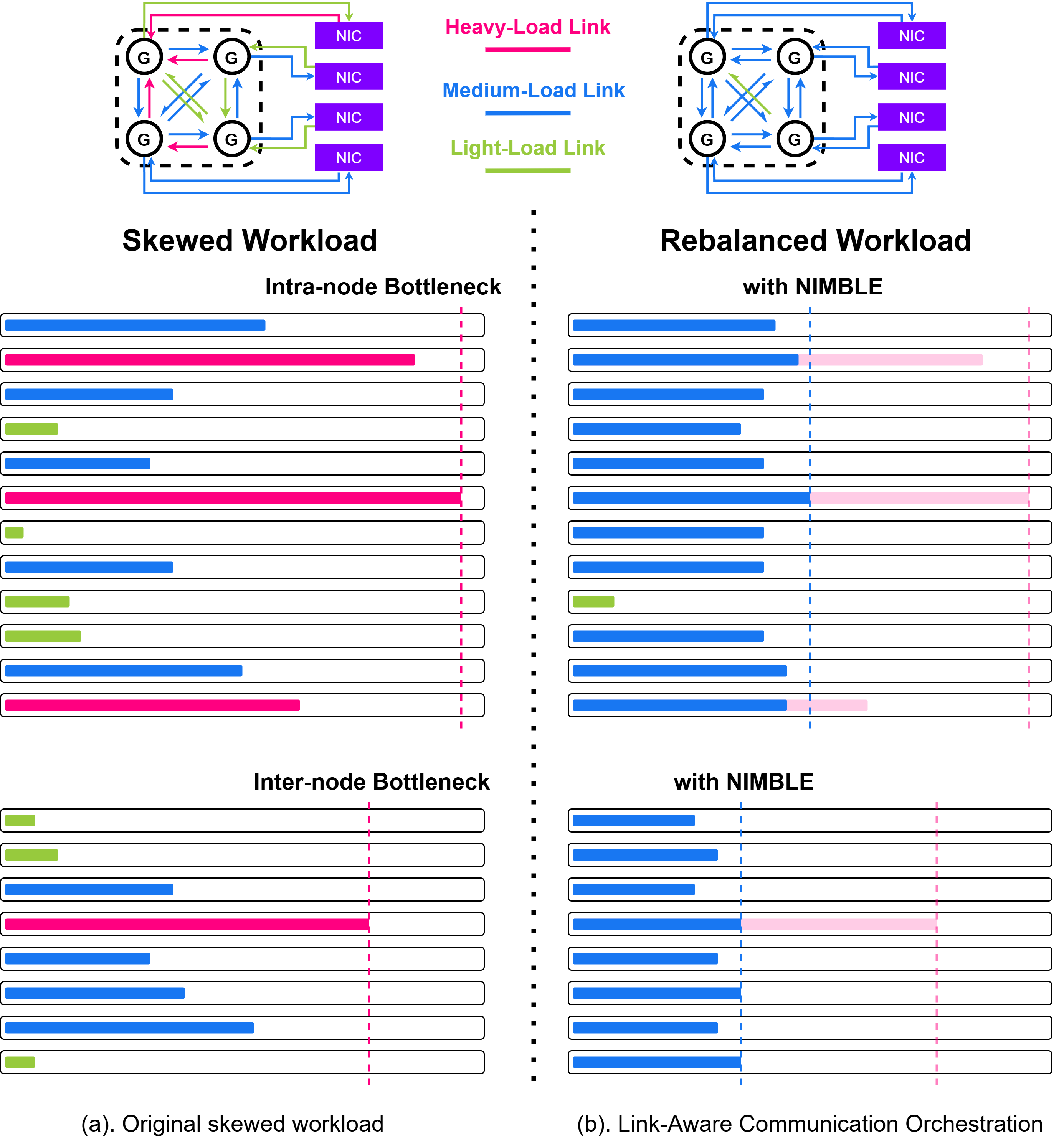}
    \caption{In skewed workload, heavy-load links (both intra-node and inter-node fabrics) dominate the overall latency. Our proposed NIMBLE is able to evaluate load imbalance and automatically redirect message transfer through less busy paths, reducing the overall maximum latency across all links. }
    \label{fig:draft}
\end{figure}

However, practical application workloads frequently exhibit non-uniform, dynamically evolving, or inherently sparse communication patterns that prevent them from achieving balanced utilization across available links. Commonly observed scenarios include skewed All-to-Allv exchanges in Mixture-of-Experts (MoE) in recent large language models (LLMs) \cite{fedus2022switch,rajbhandari2022deepspeedmoe,hwang2023tutel}, irregular communication in adaptive numerical simulations, bursty traffic in recommendation and inference pipelines \cite{naumov2019dlrm,zhong2024distserve}, and uneven data distribution in distributed graph and sparse algebra computations \cite{malewicz2010pregel,bell2009spmv}. In such cases, certain links become saturated—leading to congestion hotspots—while others remain underutilized or completely idle. The consequence is a substantial degradation in effective throughput, significant increase in tail latencies (p99), and compromised overall scalability, even though the total communication volume is well within the theoretical bandwidth capacity of the system.

Existing communication middleware libraries, such as NVIDIA NCCL, AMD RCCL, and MPI combined with UCX, largely rely on static strategies. For instance, they typically establish ring or tree topologies based on fastest-path measurements at initialization time or employ simplistic multi-rail hashing to spread traffic across available NIC rails \cite{nvidia-nccl-pxn,ucx-faq,ucx-hpcx-doc,liu2004multirail}. These static schemes, while effective for balanced traffic, are fundamentally unable to respond dynamically when traffic patterns deviate from initial assumptions. As a result, load imbalances persist or worsen, creating chronic bottlenecks at the same few links repeatedly across iterations or phases, causing systematic inefficiencies.

To address this fundamental gap, we present \textbf{NIMBLE}, a novel runtime framework that dynamically orchestrates traffic distribution across intra-node and inter-node links to minimize communication bottlenecks. Unlike previous approaches, NIMBLE is fundamentally \textit{endpoint-driven}, capable of making real-time traffic distribution decisions from the perspective of the communicating endpoints (GPUs/NICs), and explicitly exploits intra-node GPU-to-GPU connectivity in conjunction with multi-rail inter-node networks. NIMBLE performs its decisions using a capacity-normalized optimization formulation designed specifically to minimize the maximum congestion across all available links, employing a bottleneck-aware path-cost metric that mirrors the pipeline throughput characteristics of real-world GPU-to-GPU and GPU-to-NIC paths \cite{garg1998faster,madry2010multiflow}.

In addition to its novel control-plane optimization, NIMBLE incorporates a highly efficient GPU kernel-based RDMA pipelining mechanism that allows data transfers to traverse multiple GPU hops and NIC rails transparently and concurrently, ensuring minimal overhead and maximum link saturation \cite{cuda-gpudirect-rdma,nvidia-nccl-pxn}. To guarantee practicality and correctness, NIMBLE employs carefully designed policies: a size threshold that prevents excessive fragmentation of small messages, per-destination reassembly queues to maintain ordering semantics, and hysteresis-based load metrics to avoid oscillations in path selection.

In summary, the key contributions of our work are:
\begin{itemize}
    \item \textbf{Problem identification and quantification.} We identify communication imbalance as a critical runtime bottleneck across a wide spectrum of HPC and AI workloads and systematically quantify its impact on throughput and tail latency.
    \item \textbf{Runtime orchestration.} We design and implement \textbf{NIMBLE}, an endpoint-driven framework that continuously rebalances communication load by dynamically splitting and redistributing traffic across intra-node and inter-node links, while preserving ordering and determinism.
    \item \textbf{Optimization and fast dataplane.} We introduce a capacity-normalized, bottleneck-aware optimization solved via a multiplicative-weights method and pair it with a GPU kernel-based RDMA pipelining engine that enables efficient multi-hop forwarding across intermediate GPUs and rail-matched NICs. Practical policies (size thresholds, per-destination reassembly, hysteresis) avoid fragmentation and oscillation.
    \item \textbf{Evaluation highlights.} On H100-SXM5 nodes with NVLink 4 and four NDR400 rails per node, NIMBLE:
    \begin{itemize}
        \item \emph{Intra-node:} boosts peak GPU$\rightarrow$GPU bandwidth from 120~GB/s (direct) to 213.1~GB/s with one intermediate hop path and 278.2~GB/s with two additional paths; saturation occurs beyond $\sim$64~MB, and multi-pathing is disabled for $\le$1~MB to avoid overhead.
        \item \emph{Inter-node:} achieves 170.0~GB/s aggregate across four rails (single-rail 45.1~GB/s) with minimal forwarding overhead when enforcing rail matching.
        \item \emph{Skewed All-to-Allv (MoE, 8 experts, 2 nodes):} improves token dispatch and combine throughput by up to \textbf{5.2$\times$} over NCCL; NIMBLE matches the baseline in mild skew/small-message regimes where both kernel-based and OpenMPI’s DMA-driven path may slightly outperform. On end-to-end LLM MoE benchmark, NIMBLE achieves up to \textbf{1.35$\times$} speedup.
        \item \emph{Asynchronous \emph{send/recv} (point-to-point):} provides \textbf{1.15–2.3$\times$} speedup at 8~MB and up to \textbf{3.4$\times$} at 256~MB over the baseline as imbalance grows, while matching baselines under balanced traffic.
    \end{itemize}

\end{itemize}

We believe \textbf{NIMBLE} represents a significant step toward more flexible, scalable, and efficient communication orchestration in heterogeneous GPU clusters, opening new possibilities for application developers to design less constrained, more communication-intensive algorithms and models.

%% file: content/2-background_motivation.tex
\section{Background and Motivation}

\subsection{Heterogeneous GPU Communication Fabrics}
Modern GPU clusters combine high-bandwidth \emph{intra-node} fabrics—e.g., NVIDIA NVLink/NVSwitch and AMD Infinity Fabric/xGMI—with \emph{inter-node} networks such as multi-rail InfiniBand NDR and RoCEv2. Fourth-generation NVLink exposes tens of GB/s per link (raw) with many links per GPU and NVSwitch backplanes to create dense one- or two-hop connectivity among devices \cite{nvidia-nvlink,nvidia-nvlink4-nvswitch}. AMD’s Instinct MI300X platform similarly provides multi-hundreds of GB/s aggregate xGMI bandwidth across the eight-GPU complex \cite{amd-mi300x-datasheet,amd-mi300x-rccl-xgmi}. At the rack/fabric level, NDR (400\,Gb/s per port) and HDR (200\,Gb/s) InfiniBand provide low-latency, full-duplex links with multiple NICs (rails) per host \cite{nvidia-ndr-overview,mellanox-qm97x-specs}. RoCEv2 enables routable RDMA over IP, frequently deployed alongside congestion control (e.g., DCQCN/HPCC) in multi-tenant clusters \cite{nvidia-rocev2,cisco-rocev2}. GPU–NIC affinity and GPUDirect RDMA allow direct DMA between GPUs and HCAs, avoiding host staging and exposing the full path diversity within and across nodes \cite{cuda-gpudirect-rdma,gpu-operator-rdma}.

\subsection{Limitations of Current Communication Libraries}
State-of-practice libraries (NCCL/RCCL; MPI over UCX) are highly optimized but largely \emph{static} in how they construct paths. NCCL discovers device/topology at initialization and builds rings/trees (and since v2.12, offers PXN to better exploit rail-local layouts for \texttt{alltoall}) \cite{nvidia-nccl-pxn,nccl-issue-918,nccl-issue-rail}. UCX provides cross-transport \emph{multi-rail} striping at the transport layer but does not (by itself) re-slice per-message GPU traffic using live link pressure \cite{ucx-faq,ucx-hpcx-doc}. Classic MPI multi-rail work demonstrates bandwidth gains by striping across multiple HCAs but remains a flow-level technique rather than an endpoint-level, runtime path orchestrator \cite{liu2004multirail}. In practice, when traffic skews at runtime, these static or transport-granularity schemes can leave some NVLink edges and NIC rails idle while others saturate—motivating an endpoint runtime that adapts path usage online.

%% file: content/3-Overview_of_imbalanced.tex
\section{Overview of Communication Imbalance}

In this section, we formally characterize communication imbalance, systematically classify common imbalance scenarios encountered in high-performance computing (HPC) and artificial intelligence (AI) workloads, and clearly define the metrics used to quantify imbalance and evaluate the effectiveness of runtime orchestration solutions.

\subsection{Classification of Communication Imbalance Patterns}

Communication imbalance is not monolithic; it emerges in distinct patterns across various application domains and scenarios. We categorize representative imbalance patterns observed in practice:

\paragraph{ Skewed All-to-All(v) Communication}  
This imbalance pattern commonly arises in distributed machine learning models employing expert parallelism, such as Mixture-of-Experts (MoE) architectures, where tokens or inputs are dynamically routed among experts hosted across distributed GPUs. Runtime data-dependent routing decisions can create severe skew, directing a disproportionate number of messages to a subset of GPUs or nodes, resulting in overloaded links.

\paragraph{ Sparse Many-to-Few (Aggregator) Communication}  
Frequently observed in distributed reductions, parameter servers, or aggregation services, this pattern involves numerous sources sending data simultaneously to a limited number of aggregation destinations, causing severe imbalance at destination-bound links.

\paragraph{ Neighbor-Exchange with Boundary Hotspots (Stencil Computations)}  
In HPC applications such as stencil computations, adaptive mesh refinement, or particle simulations, ranks at domain boundaries may have disproportionately heavy communication with neighbor ranks. Such imbalance creates hotspots at specific intra-node or inter-node boundaries.

\paragraph{ Irregular Point-to-Point Patterns}  
Graph algorithms (e.g., BFS, PageRank), sparse numerical computations (e.g., SpMV), and adaptive scientific applications often generate irregular communication patterns. These workloads inherently lead to uneven link utilization depending on input data characteristics.

\subsection{Why Existing Approaches Fail}

Current communication libraries and middleware—such as NCCL, RCCL, and MPI/UCX—rely on static routing mechanisms. NCCL employs precomputed ring or tree topologies optimized at initialization time. MPI/UCX multi-rail configurations typically use simple static hashing or round-robin techniques for rail selection. While such approaches are effective under uniform workloads, they fundamentally lack the ability to respond dynamically when imbalance emerges, thus failing to redistribute traffic away from congestion hotspots effectively. Static routing exacerbates imbalance by consistently funneling data through the same links, increasing tail latency and causing systemic under-utilization.

\subsection{Opportunity for Dynamic Endpoint-Orchestrated Balancing}

The above analysis motivates our approach: a runtime solution positioned directly at the communication endpoints (GPUs and NICs), capable of dynamically observing real-time load conditions and redistributing communication across all available intra-node and inter-node paths. The key insight is that imbalance can be resolved proactively at the endpoints before congestion saturates the network fabric. Achieving this dynamic adaptation requires:

\begin{enumerate}
    \item Runtime on-the-fly per-link utilization balancing.
    \item Rapid optimization of path assignments based on a capacity-normalized, bottleneck-aware metric.
    \item Efficient kernel-based pipelining for data forwarding along multi-hop paths through intermediate GPUs and rail-matched NICs, with minimal overhead.
\end{enumerate}

In the following sections, we describe the design, implementation, and detailed evaluation of \textbf{NIMBLE}, a system embodying precisely this vision, demonstrating that endpoint-driven dynamic orchestration significantly mitigates imbalance, boosts aggregate throughput toward the sum of available capacities, and substantially reduces tail communication latencies.

%% file: content/4-design.tex
\section{Design of NIMBLE}
\label{sec:design}
\subsection{Overview}

NIMBLE (\textit{Node-Interconnect Multi-path BaLancing with Execution-time orchestration}) is a runtime framework designed to mitigate communication skew in modern HPC systems by dynamically redistributing traffic across all available interconnects. It enables adaptive, fine-grained communication scattering without requiring any modifications to application logic.

\begin{figure}[h]
    \centering
    \includegraphics[width=0.40\textwidth]{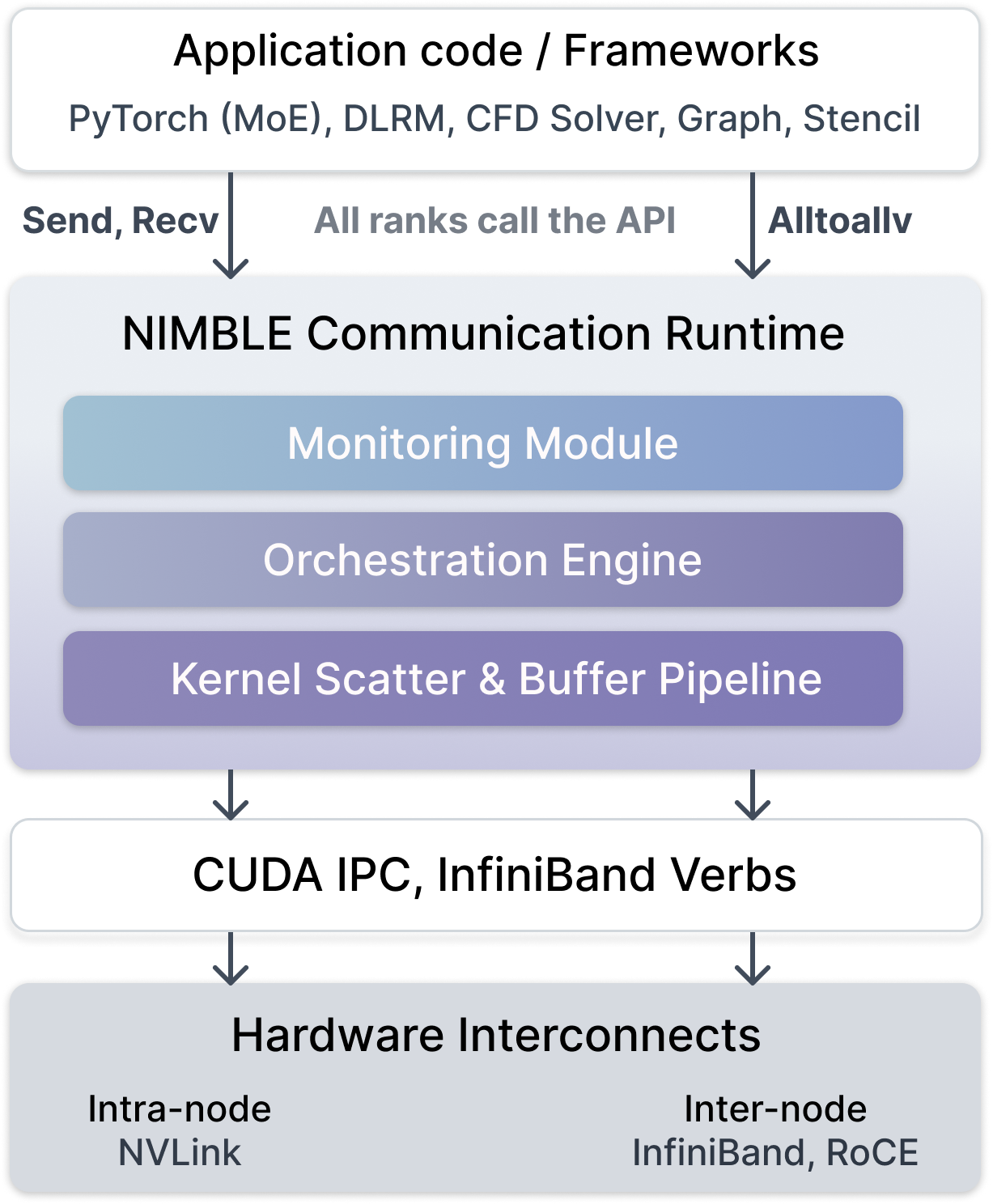}
    \caption{NIMBLE system architecture. Positioned between applications and hardware fabrics, NIMBLE intercepts communication calls and orchestrates dynamic link-aware data transfers across intra- and inter-node interconnects.}
    \label{fig:laco}
\end{figure}

Figure~\ref{fig:laco} illustrates the architecture of NIMBLE. The runtime consists of three key components: 
(1) a lightweight monitoring module that captures runtime communication patterns, 
(2) an orchestration engine that detects skew and determines link-aware redistribution strategies based on real-time path utilization, and 
(3) a \textit{Kernel Scatter \& Buffer Pipeline} that maps GPU thread blocks to specific links and enables pipelined, nonblocking data transfers across multiple intermediate GPUs and NICs simultaneously. The buffer pipeline ensures that intermediate buffers are aligned and overlapped in a way that allows continuous forwarding through multiple paths without serialization or host-side intervention.

Together, these components enable NIMBLE to operate transparently within existing HPC and AI software stacks while exploiting hardware-level bandwidth diversity. By treating communication links as orchestratable and schedulable resources, NIMBLE provides a general, robust solution to the growing problem of skewed communication in large-scale, irregular workloads.

\subsection{Hop-Adaptive MCF load balancing}
At the heart of NIMBLE’s orchestration capability lies a dynamic load-balancing strategy that adapts communication flows to current network conditions. Rather than relying on static, shortest-path routing as done in conventional libraries (e.g., NCCL, MPI), NIMBLE actively monitors link utilization and redistributes traffic to mitigate emerging bottlenecks. This subsection details our formulation of communication load balancing as a link-aware routing problem over the system’s interconnect topology, enabling NIMBLE to reshape communication patterns in real time for improved throughput and reduced contention.

\begin{figure}[h]
    \centering
    \includegraphics[width=0.47\textwidth]{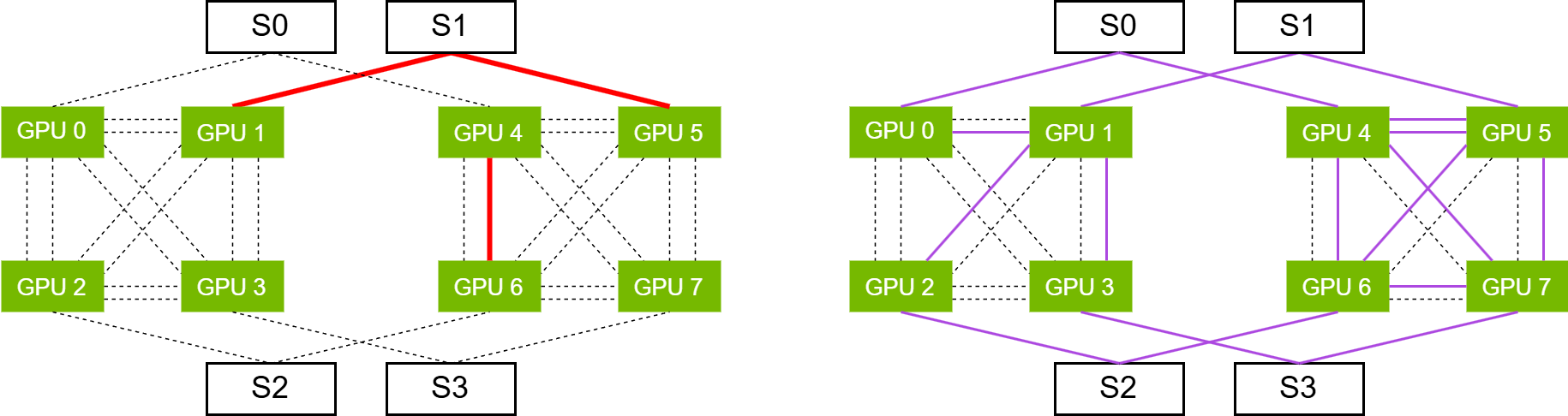}
    \caption{An example of NIMBLE orchestrating communication workload from hotspot links to all available links in the system. We significantly increased the overall throughput by scattering the original point-to-point workload into an aggregation of communication paths.}
    \label{fig:congested_link}
\end{figure}

Given a skewed workload and the interconnect topology illustrated in Figure~\ref{fig:congested_link}, our goal is to identify the links with potential congestion and reroute portions of the workload to alternate paths that are less loaded or idle.

Current communication libraries such as NCCL, MPI, MSCCL, etc., perform the communication in a least-hop/fastest-path manner. During the initialization phase, they will detect the hardware topology, i.e. NVLink, PCIe, Networks, then for every pair of devices, they will identify the fastest peer-to-peer connection path. For example, on systems equipped with NVLink (or similar fabrics), GPUs on the same node will always use the direct-connected NVLink for point-to-point communication. For GPUs on different nodes, the fastest path will include the NICs that have the best hardware affinity to the GPUs. Some libraries such as NCCL also favor railed-matched NICs to reduce switch-level hops and hence, reduce latency. However, as shown in above figure, these solutions can lead to severe congestion on links on the path of heavily loaded peers, wasting available bandwidth on other links.

Conceptually, we aim to find a routing scheme that minimizes the maximum load across all links, which can be formulated precisely as an Integer Programming (IP) problem. Let $K$ represent the set of traffic demands (message), $E$ the set of directed links (edges), and $V$ the set of nodes (GPUs). We can define integer variables $x_{k,e}$ representing the amount of flow for message $k \in K$ traversing link $e \in E$, and an integer variable $Z$ representing the maximum load objective. The optimization problem seeks to:
\begin{equation}
    \text{Minimize } Z
    \label{eq:ip_objective}
\end{equation}
Subject to:
\begin{align}
    \sum_{e \in \text{out}(v)} x_{k,e} - \sum_{e \in \text{in}(v)} x_{k,e} &=
    \begin{cases} 
        d_k & \text{if } v = s_k \\
        -d_k & \text{if } v = t_k \\
        0 & \text{otherwise}
    \end{cases} \quad \forall k \in K, v \in V \\
    \sum_{k \in K} x_{k,e} &\le Z \quad \forall e \in E \label{eq:ip_maxload} \\
    x_{k,e} &\in \mathbb{Z}_{\ge 0} \quad \forall k \in K, e \in E \label{eq:ip_integrality} \\
    Z &\in \mathbb{Z}_{\ge 0} \label{eq:ip_z_integrality}
\end{align}

\input{algorithm/gk_intra}

where $d_k$ is the demand for message $k=(s_k, t_k)$, $\text{out}(v)$ are links originating at $v$, and $\text{in}(v)$ are links terminating at $v$. While exact solutions can be obtained using specialized IP solvers, solving such integer multi-commodity flow (MCF) problems is generally NP-hard. The computational complexity grows rapidly with the number of nodes, links, and messages. For dynamic network environments where routing decisions must be made rapidly -- potentially within timeframes comparable to the communication latency itself (milliseconds to seconds) -- the processing time required by IP solvers (which can range from minutes to hours or more for non-trivial instances) renders this approach infeasible for runtime execution.

Therefore, to achieve efficient load balancing within practical time constraints, we employ an iterative approximation algorithm. The approach is inspired by the multiplicative weights update (MWU) framework, notably utilized in the Garg-Könemann algorithm for minimum congestion multi-commodity flow~\cite{garg2007faster}.

\begin{figure}[t]
    \centering
    \includegraphics[width=0.47\textwidth]{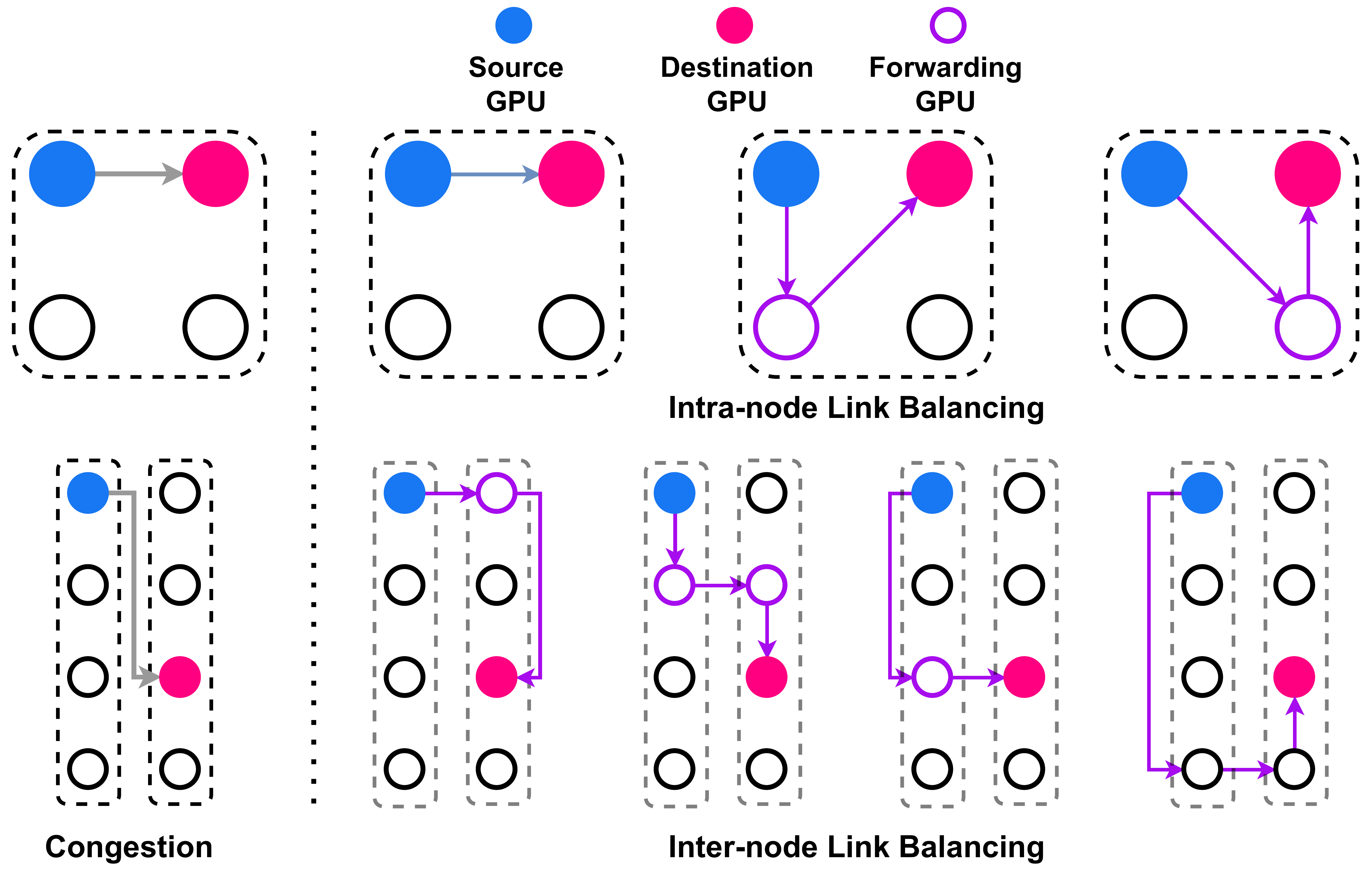}
    \caption{Given 4-GPU 4-NIC computing nodes, intra-node link balancing tries to utilize all available links that direct the path from sources GPU to destination GPU; Inter-node link balancing leverages the rail-matched path among nodes as well as intra-node links on both nodes.}
    \label{fig:rebalancing}
    \vspace{-1em}
\end{figure}

\textbf{} The core strategy involves adaptively routing flow increments in successive iterations. Communication link costs are dynamically updated based on their current load to penalize congestion. While the original Garg-Könemann algorithm typically employs an exponential cost function, e.g., $c_e \propto \exp(\alpha L_e)$ \cite{garg2007faster}, our implementation utilizes a custom cost function $c_e = \mathcal{F}(L_e)$. This function $\mathcal{F}$ is specifically designed according to hardware features and potential overhead in multi-path routing (detailed discussed in Section~\ref{sec:fundamental_exp}) while still ensuring that link costs increase sharply with load $L_e$, thereby discouraging the overuse of congested links. Note that the path cost is usually calculated as the standard sum of individual link costs along the path ($\sum_{e \in P} c_e$), which is fundamentally applied to shortest-path algorithms like Dijkstra and ensures that the cumulative impact of congestion along the entire path informs the routing decision. However, here we take the \textbf{maximum} link cost on the path, the reason is that (as will discuss later in section~\ref{sec:pipelining_rdma}) in our implementation, we adopt fine-grained pipelined buffers for GPU RDMA in moving data between GPUs and NICs along the path, forming a non-blocking streamline, where the overall bandwidth is mostly affected by the bottleneck link in the path.

In order to schedule communication jobs to the corresponding links and make sure all data are properly transferred, we loop over all rank pairs. For pairs that have actual workload, we apply an iterative assigning strategy. Each time we examine a given pair $(s, d)$, where $s$ denotes the source rank and $d$ denotes the destination rank, we take a fraction of its remaining workload, denoted as $\lambda$, and try to route it through the currently least congested path. In this manner, as we come to the same pair $(s, d)$ multiple times, we are dealing with a proportionally smaller workload, i.e. $(1-\lambda)^n$ of the original workload. This can lead to a better approximation and convergence to the global optimal routing solution~\cite{garg2007faster}.

For intra-node fabrics and inter-node connections, we design dedicated strategies. First, if the source rank and destination rank are on the same node, then we can either 1). directly route the data, which will consume the bandwidth of link $e_{s,d}$, or 2) take a third rank $i$ (also in the same node) as an intermediate hop, which will consume the bandwidth of both link $e_{s,i}$ and $e_{i,d}$. The selection of the least congested path relies on the routing cost of each path. Specifically, the cost is calculated as:
\begin{itemize}
    \item \textbf{Intra-node direct path}: $c_{s,d}$
    \item \textbf{Intra-node 2-Hop path}: $max(c_{s,i}, c_{i,d})$
    \item \textbf{Inter-node path}: $max(c_{s,i_s}, c_{i_s,i_d}, c_{i_d, d})$
\end{itemize}

For intra-node direct path, it is straightforward that we use the cost of the direct link. For intra-node path with intermediate ranks, we only consider 1 additional hop, as the rest of GPUs can be part of more potential paths. For inter-node path, we need to carefully examine. In modern HPC clusters, a single computing node is often equipped with multiple NICs and interconnect fabrics. With GPU RDMA, the hardware affinity between NIC and GPU is crucial for reducing latency and achieving higher bandwidth. A typical matching for a computing node with same number of GPUs and NICs is to let each GPU attach to the NIC according to their ordinal index. For example, NIC 0 only performs RDMA operation with GPU 0, NIC 1 only perform RDMA operation with GPU 1, etc. 

For inter-node paths, another noteworthy hardware feature is the switch-level routing topology. In these multi-rail HPC clusters that are equipped with multiple NICs per node, the switch can also be hierarchical. Data exchanging with the same switch is often fast and of low latency. In existing advanced communication libraries such as NCCL, a unique technique called "PXN" is applied to avoid the above-mentioned rail mismatch.

\begin{table}[htbp]
    \centering
    \caption{Intra-node vs Inter-node communication latency and NIMBLE orchestration algorithm time}
    \label{tab:intra_inter_singlecol}
    \begin{tabular}{@{} c | r r | r r @{}}
        \toprule
        \textbf{Size (MB)} & 
        \multicolumn{2}{c|}{\textbf{Intra-node}} & 
        \multicolumn{2}{c}{\textbf{Inter-node}} \\
        \cmidrule(r){2-3} \cmidrule(l){4-5}
        \small \textbf{} & 
        \small \textbf{Algo (ms)} & \small \textbf{Comm (ms)} & 
        \small \textbf{Algo (ms)} & \small \textbf{Comm (ms)} \\
        \midrule
        16  & 0.0321 & 0.1973 & 0.0374 & 0.4860 \\
        32  & 0.0344 & 0.3662 & 0.0325 & 0.7361 \\
        64  & 0.0347 & 0.6512 & 0.0350 & 1.7701 \\
        128 & 0.0343 & 1.0689 & 0.0426 & 3.2270 \\
        256 & 0.0363 & 2.0464 & 0.0480 & 6.5390 \\
        \bottomrule
    \end{tabular}
    \vspace{-1em}
\end{table}

In our design, we apply the same methodology. When choosing possible paths between two nodes, we only consider rail-matched paths, i.e. using the direct connection fabric between NIC $i_s$ on the source node and NIC $i_d$ on the destination node. With this, we can guarantee to utilize all available inter-node bandwidth while also at their lowest latency. Figure~\ref{fig:rebalancing} illustrates our design in detail. On a 4-GPU 4-NIC computing node, for each GPU, it has 3 direct links to other 3 GPUs, thus, we can at most distribute the original intra-node workload to 3 different paths. For inter-node, as 4 NICs have exclusive interconnect fabrics, we can leverage all of them to achieve a theoretical 4x speedup in overall communication throughput.

Table~\ref{tab:intra_inter_singlecol} compares the overhead of our proposed link-aware workload orchestration algorithm and the overall communication latency. We use a 1D stencil as the application, where each rank communicates with its neighbors. As the message size goes larger, we are observing a slightly increasing trend in the algorithm overhead, e.g., from 0.0321~ms to 0.0363~ms in intra-node, and 0.0374~ms to 0.0480~ms in inter-node. However, they are too negligible compared to the actual communication.

\subsection{Pipelining kernel-based GPU RDMA}

\begin{figure}[h]
    \centering
    \includegraphics[width=0.4\textwidth]{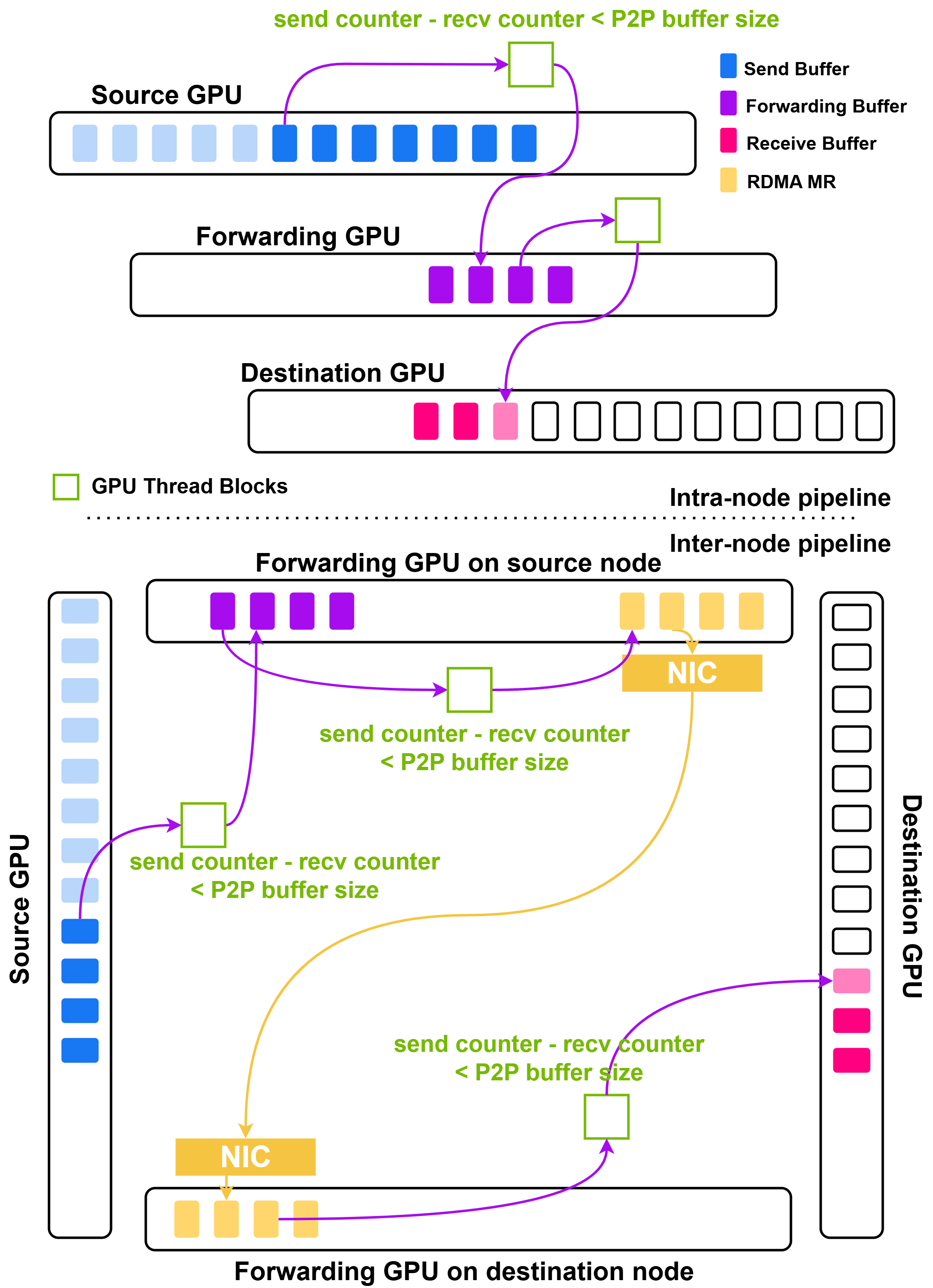}
    \caption{On every forwarding GPU, we allocate a small amount of peer-to-peer buffer and register RDMA memory region if NIC is involved in the path. These intermediate buffers are much smaller than the actual data buffer on the sender and receiver side, they serve as the buffer pool in the pipeline.}
    \label{fig:buffers}
\end{figure}
\label{sec:pipelining_rdma}

Efficient inter-GPU communication is pivotal in high-performance computing (HPC) environments, especially when addressing imbalanced workloads and maximizing hardware utilization. Traditional host-initiated data transfer methods, such as \texttt{cudaMemcpy} and \texttt{cudaMemcpyPeer}, involve the CPU orchestrating data movement between GPUs. While these methods are straightforward for basic point-to-point transfers, they introduce overhead due to CPU invocation, scheduling, and synchronization. In contrast, GPU kernel-based data transfer mechanisms, as implemented in libraries like NVIDIA's NCCL, delegate the control of data movement to the GPUs themselves. NCCL employs specialized kernels that execute on the GPUs to coordinate and perform data exchanges. This paradigm allows for the implementation of sophisticated communication patterns, such as ring or tree algorithms, which can decompose large transfers into optimized stages, overlap communication phases, and maximize hardware utilization \cite{nvidia_nccl, hoefler2015remote}. In our design, we follow this kernel-based scheme to create a perfect data moving pipeline across multiple hops.

Figure~\ref{fig:buffers} depicts an intra-node 2-hop GPU forward path and an inter-node 3-hop GPU-NIC forward path. On the sender side, the blue blocks denote the actual send buffer. We take the inter-node case as an example, as it includes the intra-node part. On the receiver side, pink blocks denote the actual receive buffer. On two intermediate forwarding GPUs, we only allocate a very small piece of P2P buffer. The sender GPU will launch thread blocks that move data from the send buffer to the first intermediate GPU's P2P buffer (denoted as purple blocks). These thread blocks will coordinate with the thread blocks on the peer GPU. Then, the thread blocks on the first intermediate GPU will move data from the P2P buffer to NIC's pre-allocated RDMA memory region (MR). Note that, like NCCL, in our design, the RDMA primitives (i.e. ibv\_post) will be issued by the CPU thread, therefore, these kernels need to coordinate with CPU for inter-node RDMA. Similarly on the receiver side. 

Since the intermediate P2P buffer is much smaller than the actual data size, a fine-grained synchronization mechanism is implemented to avoid data race. Each GPU has pre-allocated peer-accessible counters that can be modified and read by its peer GPUs. On the sender side, before it actually issues the send, it will check if the size of the on-the-fly data to its peer GPU (by comparing the counter of data sent and the counter of data received) has exceeded the P2P buffer size, and only when the P2P buffer has available slots, will the send GPU keep issuing send. On the intermediate forwarding GPUs, the synchronization is trickier. Each forwarding GPU needs two pairs of sent and received counters, for its sending peer and its receiving peer. The data movement will be blocked if any one pair of counters does not meet the rule we discussed above.

\subsection{Peer-exclusive kernel pairing}
In NIMBLE's orchestration, one GPU might have multiple of the following tasks running at the same time: 1) sending, 2) forwarding from one GPU to another, 3) receiving; thus, In order to make sure all assign tasks are performed in parallel, each GPU will launch kernels for each \texttt{recv} peer and \texttt{send} peer, and these thread blocks will work together. For the same \texttt{recv} peer and \texttt{send} peer, we will use the same thread blocks (channels), with a task queue of all its tasks. The reason we do not launch different groups of thread blocks (channels) for the same \texttt{send/recv} peers is that the pre-allocated P2P buffer during initialization will be persistent throughout the application, and assigning different groups of channels to the same peer will result in redundant P2P buffer allocation and introduce significant overhead at runtime. 

\subsection{Integration with existing libraries}

Our proposed NIMBLE is tailored for load-imbalanced communication scenarios, mostly Alltoallv and send/recv operations that are frequently used in AI and HPC workloads. While these communication operations are widely used, there are other collectives that are intrinsically balanced, e.g. AllReduce, Reduce-Scatter, AllGather, etc. For these collectives, NIMBLE is not involved in the routing orchestration, as existing ring, tree-based solutions already utilize all links at their maximum capacity. Thus, despite NIMBLE potentially being able to handle any communication patterns, for the consistency of use, we integrate it into the latest NVIDIA NCCL toolbox. Users can easily leverage NIMBLE with the normal NCCL \texttt{Send/Recv}, \texttt{AlltoAll} API calls.

%% file: algorithm/gk_intra.tex
\begin{algorithm}[ht]
\caption{Link Load Balancing w. Iterative Approximation}
\label{alg:intra_algo}
\begin{algorithmic}[1]
\Require GPUs $V$, NICs $N$, Links $E$, Communication pairs $K$, Workloads $d_{s,d}$, Flow fraction $\lambda$, Chunk granularity $\varepsilon$
\Ensure Path lists $Paths^{(s,d)}$, Flow amount lists $Flows^{(s,d)}$ for all $(s,d) \in K$

\State $L_e \gets 0$, $c_e \gets 0$ \textbf{for all} $e \in E$
\State $r_{s,d} \leftarrow d_{s,d}$ for all $(s, d) \in K$
\State $D_{tot} \leftarrow \sum_{(s,d)} d_{s,d}$, $R_{tot} \leftarrow D_{tot}$
\State Initialize $Paths^{(s,d)} \leftarrow \emptyset$, $Flows^{(s,d)} \leftarrow \emptyset$ for all $(s,d) \in K$ \Comment{Empty lists}

\While{$R_{tot} > 0$}
    \ForAll{pair $(s, t) \in K$ such that $r_{s,d} > 0$}
        \State $P_{best} \leftarrow \emptyset$, $c_{best} \leftarrow \infty$
        \State $e_{dir} \leftarrow (s, d)$ \Comment{\textcolor{violet}{Direct path}}
        \If{$c_{e_{dir}} < c_{best}$} $c_{best} \leftarrow c_{e_{dir}}$, $P_{best} \leftarrow \{e_{dir}\}$ \EndIf
        \ForAll{\textcolor{magenta}{$i \in V \setminus \{s, d\}$}} \Comment{\textcolor{magenta}{Intra-node 2-Hop paths}}
            \State $e_1 \leftarrow (s, i)$, $e_2 \leftarrow (i, d)$
            \State $c \leftarrow max(c_{e_1}, c_{e_2})$
            \If{$c < c_{best}$} $c_{best} \leftarrow c$, $P_{best} \leftarrow \{e_1, e_2\}$ \EndIf
        \EndFor
        \ForAll{\textcolor{teal}{$i \in N$}} \Comment{\textcolor{teal}{Inter-node NIC-included paths}}
            \State $e_1 \leftarrow (s, i_s)$, $e_2 \leftarrow (i_s, i_d)$, $e_3 \leftarrow (i_d, d)$
            \State $c \leftarrow max(c_{e_1}, c_{e_2}, c_{e_3})$
            \If{$c < c_{best}$} $c_{best} \leftarrow c$, $P_{best} \leftarrow \{e_1, e_2, e_3\}$ \EndIf
        \EndFor
        \State $f_{route} \leftarrow 0$
        \If{$r_{s,d} < \varepsilon$}
            \State $f_{route} \leftarrow r_{s,d}$ \Comment{Residual}
        \Else
            \State $f_{route} \leftarrow \lfloor r_{s,d} * \lambda \rfloor_{\varepsilon}$ \Comment{Multiple of $\varepsilon$}
        \EndIf

        \If{$flow_{route} > 0$}
            \State Append $P_{best}$ to $Paths^{(s,d)}$ 
            \State Append $f_{route}$ to $Flows^{(s,d)}$ 
            \ForAll{link $e \in P_{best}$}
                \State $L_e \leftarrow L_e + f_{route}$, $c_e \leftarrow \mathcal{F}(L_e)$
            \EndFor
            \State $r_{s,d} \leftarrow r_{s,d} - f_{route}$, $R_{tot} \leftarrow R_{tot} - f_{route}$
        \EndIf
    \EndFor 

\EndWhile 

\State \Return $Paths^{(s,d)}$, $Flows^{(s,d)}$
\end{algorithmic}
\end{algorithm}

%% file: content/5-Evaluation.tex
\section{Evaluation}
\label{sec:eval}
\subsection{Setup}
 \textbf{NIMBLE} leverages the Alltoall connected intra-node fabrics (i.e. NVLink), and the multiple NICs configuration for inter-node communication. This configuration is also widely adopted by HPC clusters and cloud providers.
 
\textbf{Hardware:} Each node is equipped with two Intel Xeon Platinum 8470 CPUs, four NVIDIA H100 (Hopper) SXM5 GPUs each with 94 GB HBM2e memory and NVIDIA NVLink 4. For the interconnect between computing nodes, there are four NDR400 Infiniband HCAs supporting GPUDirect, with each attaches to one GPU.

On the intra-node, each peer-to-peer connection consists of multiple thread blocks to fully saturate the NVLink bandwidth. When NIC is involved, a similar design applies. Note that in the inter-node case, the number of thread blocks for the intra-node link is equal to that of the inter-node link in order to match the InfiniBand throughput, achieving a bottleneck-free pipeline. We use P2P Write RDMA for all peers, and the P2P buffer is 10 MB per thread block. The number of overall thread blocks for communication is configurable at compile time; we follow the same rule as the NCCL's design.

\textbf{Software:} We compare our design with NCCL v2.26 stable branch and OpenMPI v5.0.7 with UCX v1.18.0 CUDA-Aware. We also make sure that GPU bindings and NICs are all in hardware affinity for all libraries.

\textbf{Applications:} We first illustrate the most straightforward cases where our proposed method can provide communication speedup. Then, we benchmark our method on the skewed All-to-allv and the real mixture-of-experts (MoE) blocks, as they represent the most commonly skewed communication patterns.

 \begin{figure*}[t]
    \centering
    \begin{subfigure}[b]{0.24\textwidth}
        \includegraphics[width=\linewidth]{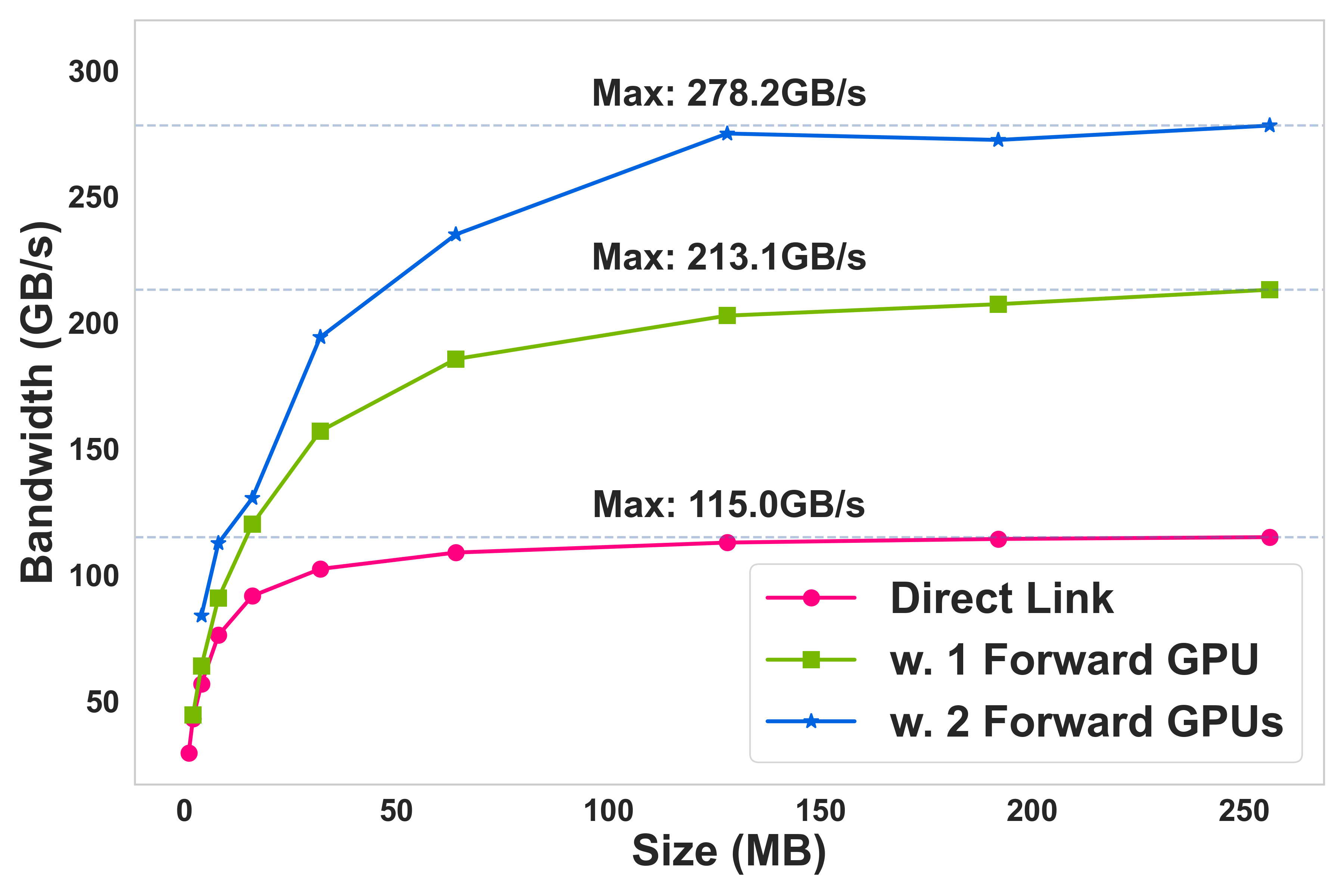}
        \caption{Intra-node multi-path aggregated bandwidth with more GPUs}
        \label{fig:sub1}
    \end{subfigure}
    \hfill
    \begin{subfigure}[b]{0.24\textwidth}
        \includegraphics[width=\linewidth]{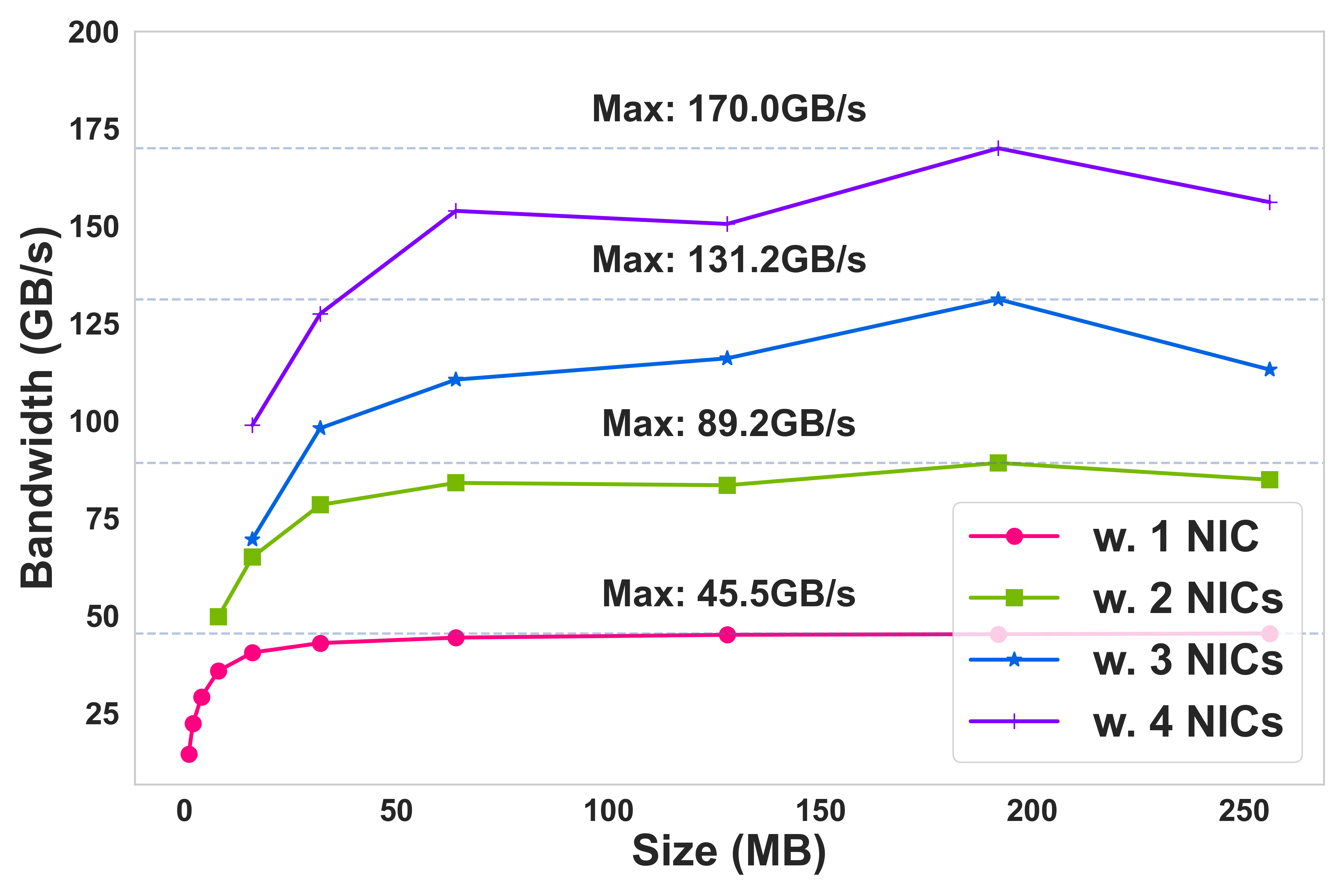}
        \caption{Inter-node multi-path aggregated bandwidth with more NICs}
        \label{fig:sub2}
    \end{subfigure}
    \hfill
    \begin{subfigure}[b]{0.24\textwidth}
        \includegraphics[width=\linewidth]{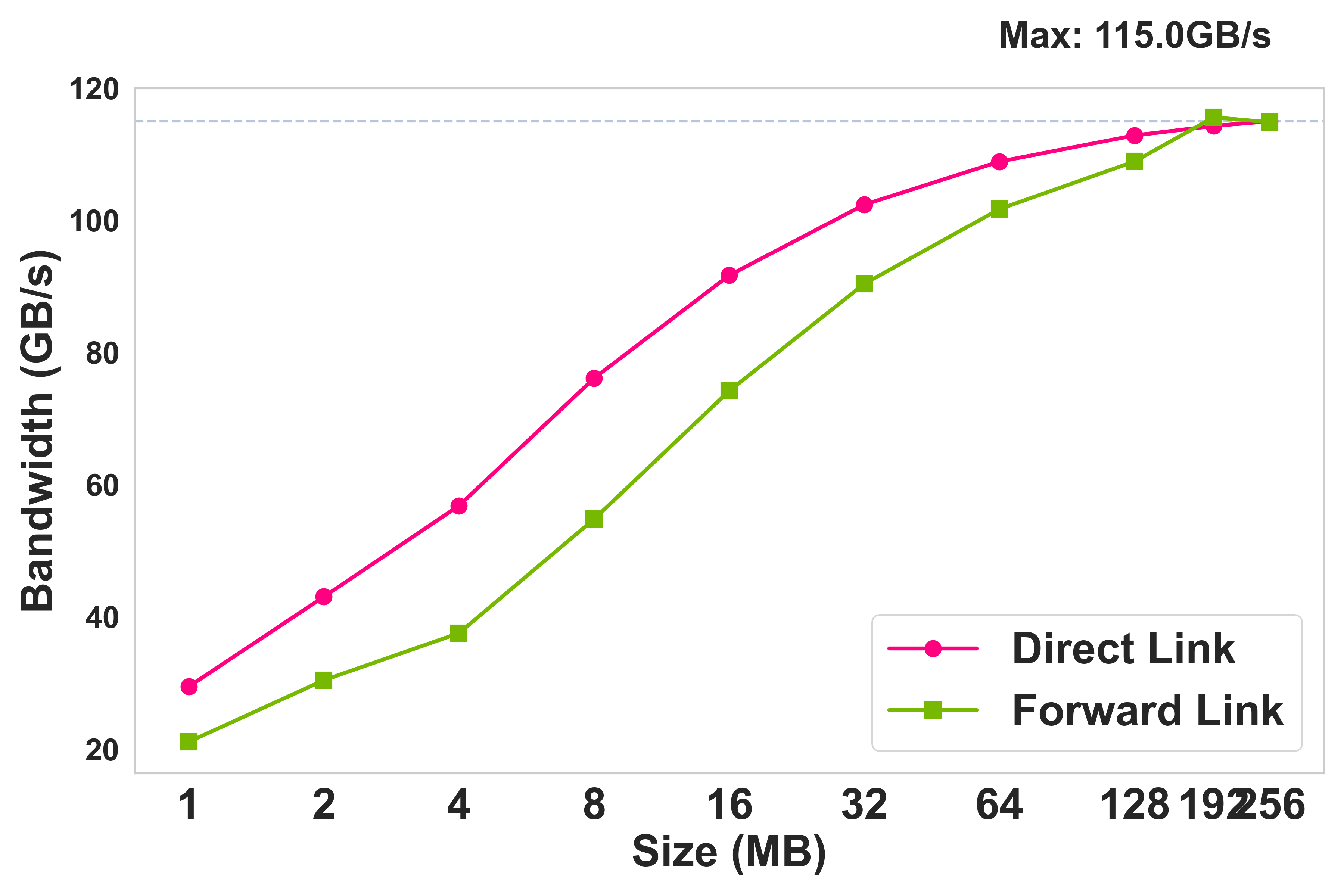}
        \caption{Intra-node standalone bandwidth of 2-hop GPU path}
        \label{fig:sub3}
    \end{subfigure}
    \hfill
    \begin{subfigure}[b]{0.24\textwidth}
        \includegraphics[width=\linewidth]{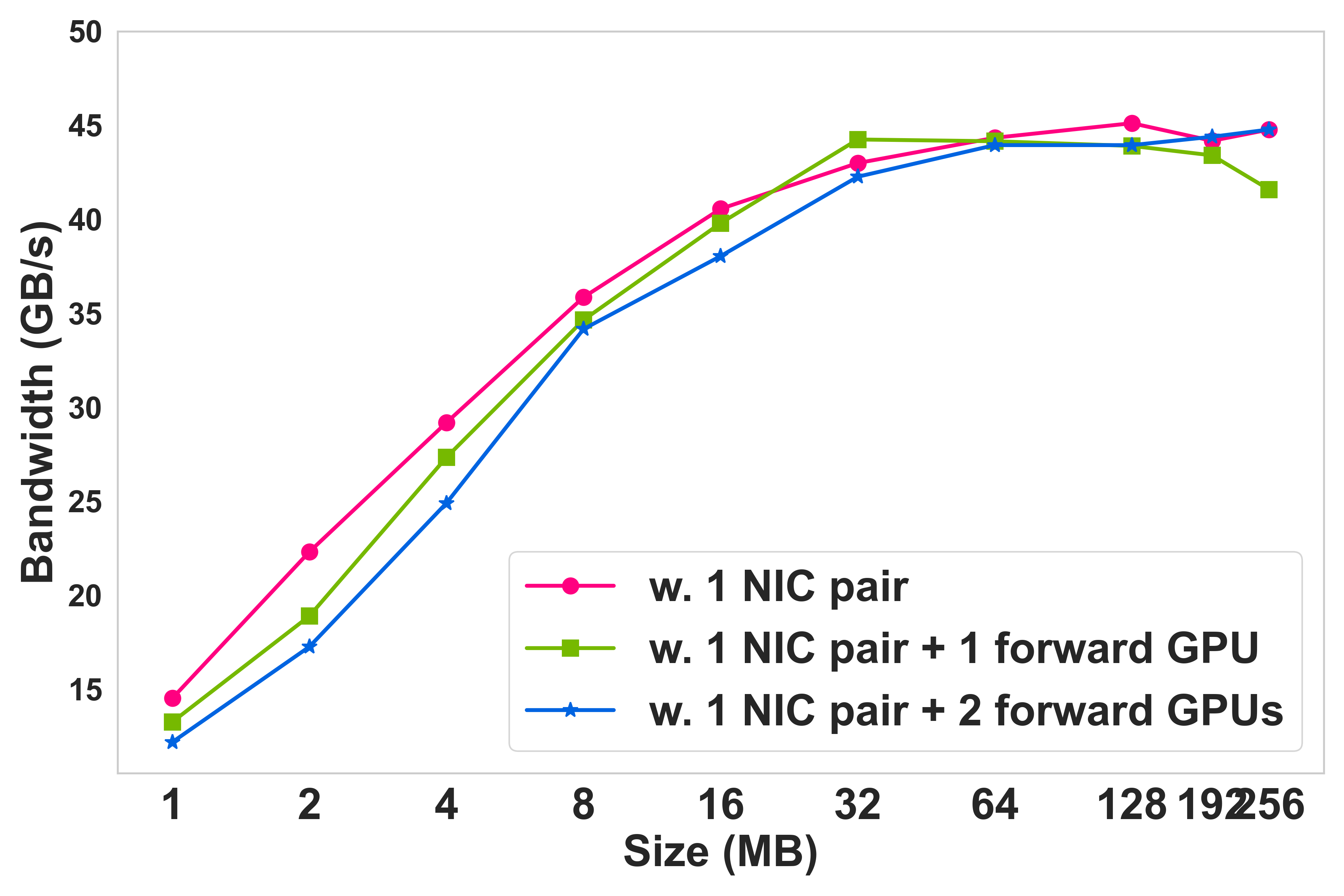}
        \caption{Inter-node standalone bandwidth of multi-hop GPU-NIC path}
        \label{fig:sub4}
    \end{subfigure}
    \caption{Demonstration of speedup from utilizing additional communication paths and the efficiency in intra-node and inter-node multi-hop paths. For intra-node, we have a theoretical speedup of 3x as we have two more idling GPUs; for inter-node, we have a theoretical speedup of 4x as we have 4 NICs in total. The number of NICs in (b) is the per-node involved NICs. The number of NICs in (d) is the rail-matched NIC pair that sits on the sender and receiver nodes.}
    \label{fig:four_figures}
\end{figure*}

\subsection{Point-to-point speedup demonstration}
\textbf{Intra-node Multi-GPU:} Within one node, GPUs are all-to-all connected via NVLink. If the direct link between sender and receiver GPUs is busy, NIMBLE uses additional GPUs as intermediates that internally invoke a "forward" operation, only transferring data without modification. Figure~\ref{fig:four_figures}(a) shows the bandwidth for intra-node communication. A direct GPU-to-GPU NVLink offers a 120 GB/s theoretical peak bandwidth, saturating around a 64 GB message size. Introducing one intermediate GPU (GPU 2) adds two additional NVLinks, reaching 213.1 GB/s peak bandwidth. Using two intermediates (GPUs 2 and 3), the configuration has six total links, achieving 278.2 GB/s peak bandwidth. Sub-linear scaling arises due to overhead in hardware message injection, potential L2 cache misses, and bandwidth contention, further discussed in Section~\ref{sec:limitations}.

\textbf{Inter-node Multi-GPU:} For inter-node transfers, multiple NICs per node can be utilized. Communication adheres to rail-matching to reduce switch latency. Our system has four NICs per node, each offering 50 Gb/s bandwidth via InfiniBand. Figure~\ref{fig:four_figures}(b) illustrates bandwidth in a scenario where one GPU per node communicates. Single NIC bandwidth saturates beyond 32 MB message size. Adding a second NIC divides the workload across two paths, nearly doubling bandwidth. Utilizing all four NICs achieves an aggregate bandwidth of 170.0 GB/s. NIC bottlenecks minimize GPU pipeline or cache-related overhead.

\textbf{Intra-node Forward Overhead:} Figure~\ref{fig:four_figures}(c) evaluates intra-node forwarding overhead against direct GPU transfers, excluding cache and HBM contention. Observed overhead primarily stems from additional pipeline setup and memory accesses on intermediate GPUs. Multi-GPU forwarding is disabled for small messages (1 MB or less) due to this significant overhead. And a significant penalty is added to the cost of routing to other links when the message size is not large enough. (refer to $\mathcal{F}$ in Algorithm~\ref{alg:intra_algo})
\label{sec:pipeline_profiling}

\textbf{Inter-node Forward Overhead:} Figure~\ref{fig:four_figures}(d) compares standalone inter-node bandwidths for rail-matched NIC paths. Direct rail-matched transfers achieve 45.1 GB/s. For rail-mismatched cases, intermediate GPUs forward data to maintain rail-matching, causing minimal overhead. NIC throughput limitations dominate performance, resulting in near-linear bandwidth gains from leveraging multiple NICs in distributed communications.

\textbf{Deeper multi-hop paths.} Our kernel pipeline generalizes to \emph{arbitrary} hop counts: by chaining per-link P2P buffers and relay kernels (Section~\ref{sec:pipelining_rdma}), NIMBLE can forward across multiple intermediate GPUs and/or rail-matched NICs without host-side serialization. However, both our measurements and the pipeline fill/flush analysis indicate diminishing—and for small/medium messages, negative—returns beyond one intra-node hop: each additional hop adds synchronization latency, increases L2/HBM traffic on relay GPUs, and lengthens the time to reach steady-state throughput. In inter-node cases where the NIC is already the bottleneck, extra GPU relays do not raise the path’s bottleneck capacity and thus yield negligible gain. Consequently, we cap candidate paths to a single intra-node hop and rail-matched inter-node routes, and encode a size-aware penalty in $\mathcal{F}(\cdot)$ (Algorithm~\ref{alg:intra_algo}) so that deeper multi-hop is only selected when message sizes are large enough to amortize pipeline overhead (consistent with Figure~\ref{fig:four_figures}).

\label{sec:fundamental_exp}
\begin{figure*}[h]
    \centering
    \begin{subfigure}[t]{0.18\textwidth}
        \includegraphics[width=\linewidth]{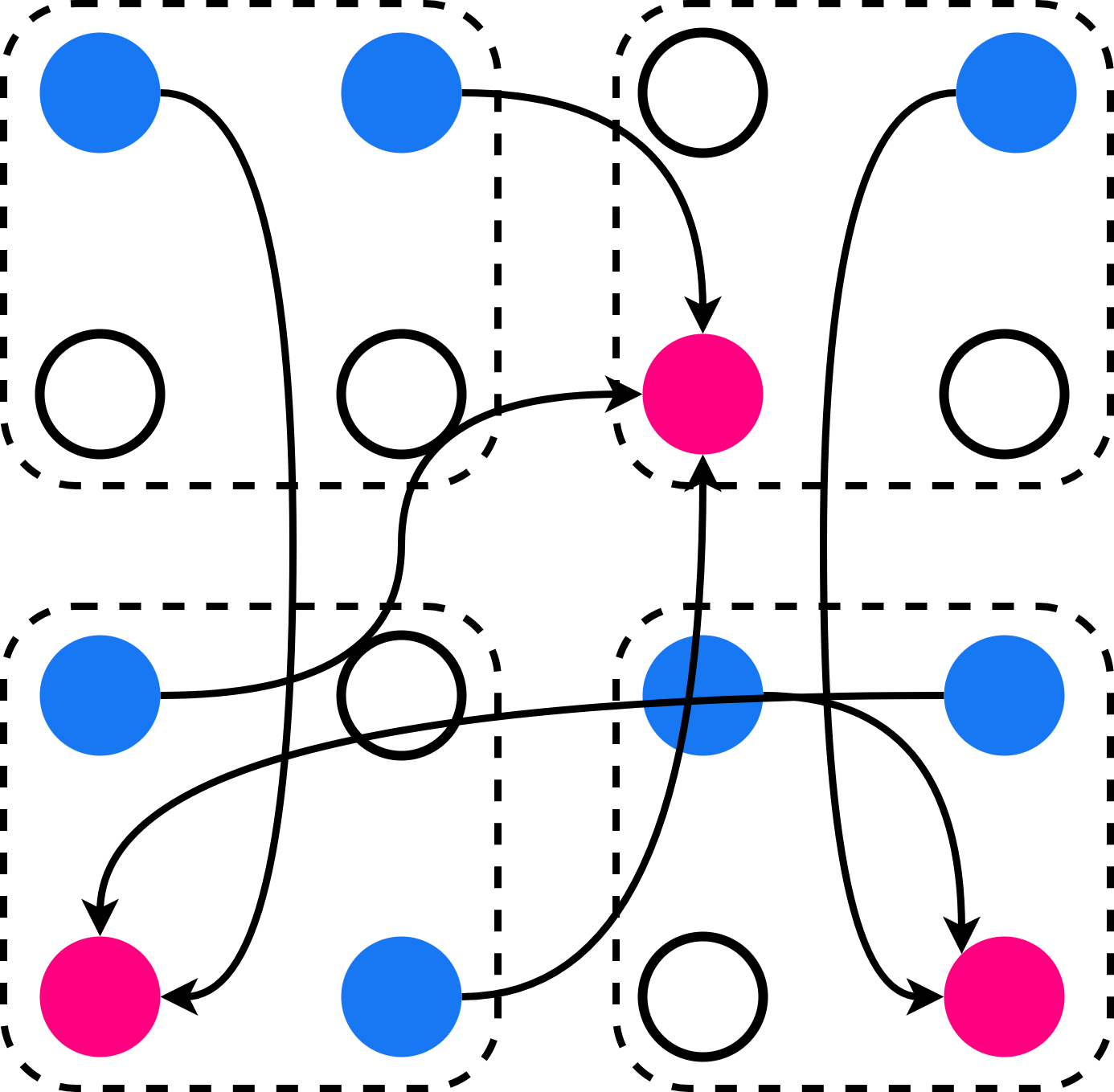}
        \caption{All-to-allv}
    \end{subfigure}%
    \hfill
    \begin{subfigure}[t]{0.24\textwidth}
        \includegraphics[width=\linewidth]{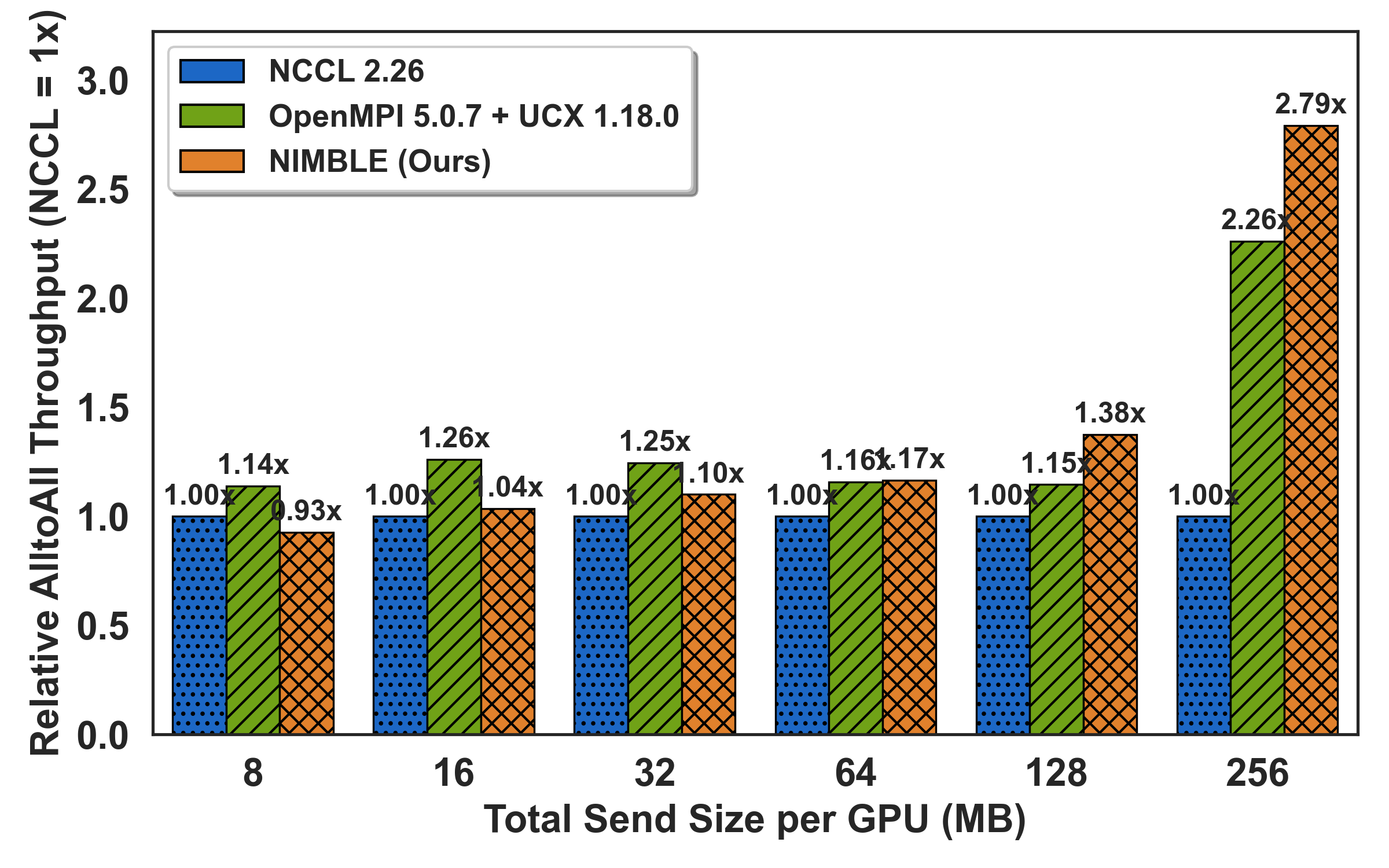}
        \caption{Hotspot ratio=0.4}
    \end{subfigure}%
    \hfill
    \begin{subfigure}[t]{0.24\textwidth}
        \includegraphics[width=\linewidth]{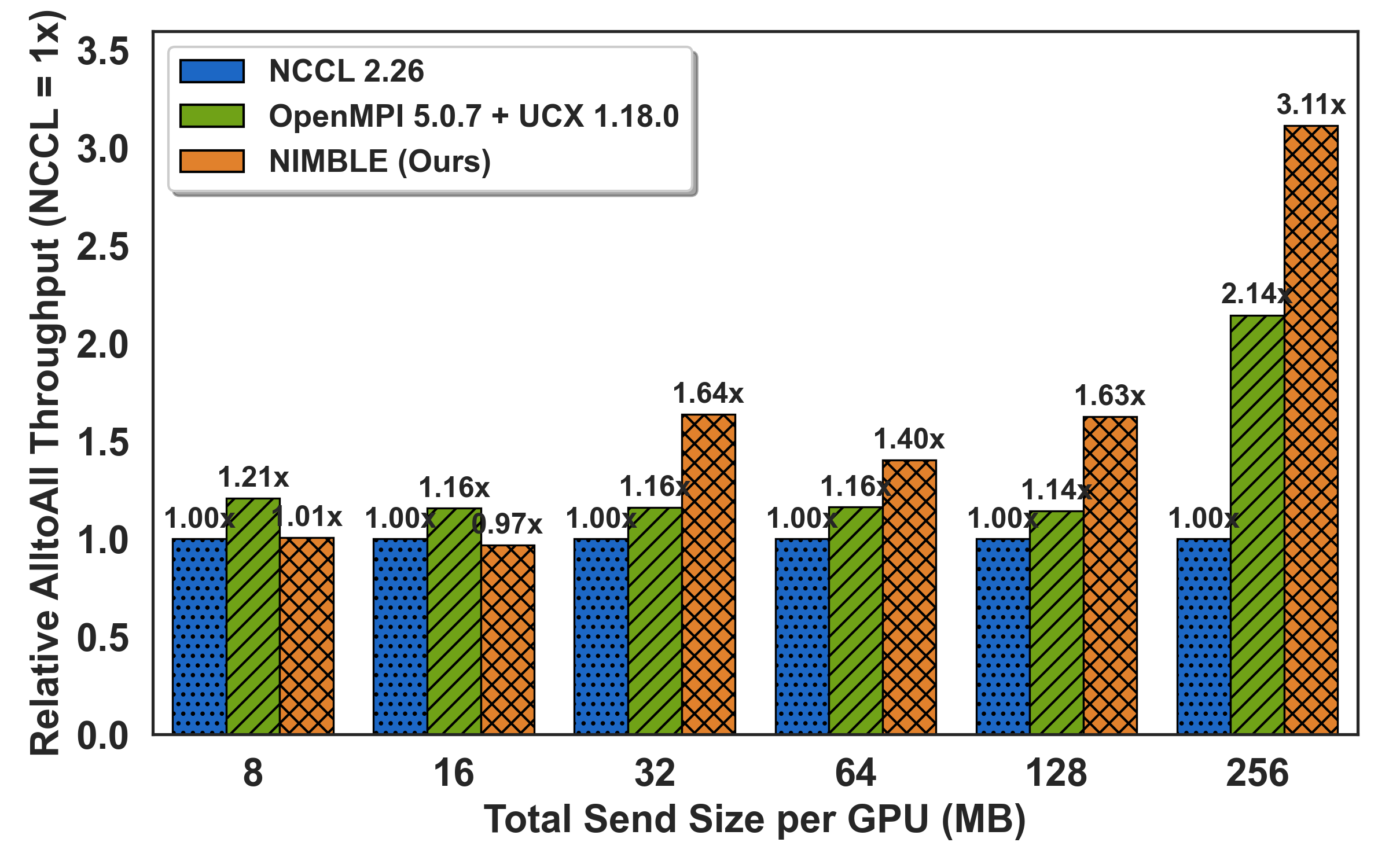}
        \caption{Hotspot ratio=0.5}
    \end{subfigure}%
    \hfill
    \begin{subfigure}[t]{0.24\textwidth}
        \includegraphics[width=\linewidth]{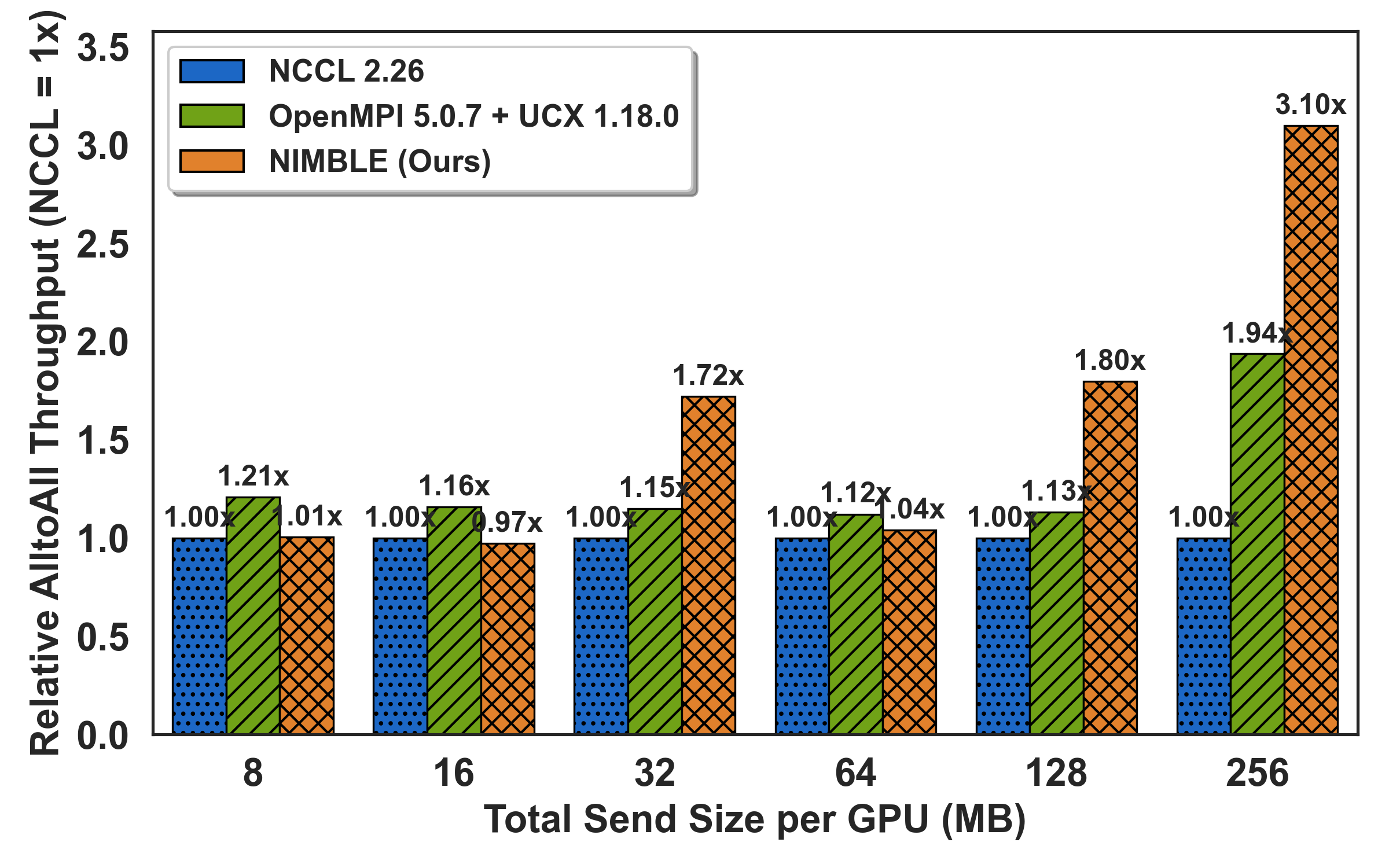}
        \caption{Hotspot ratio=0.6}
    \end{subfigure}%

    \vspace{1mm} 

    \begin{subfigure}[t]{0.24\textwidth}
        \includegraphics[width=\linewidth]{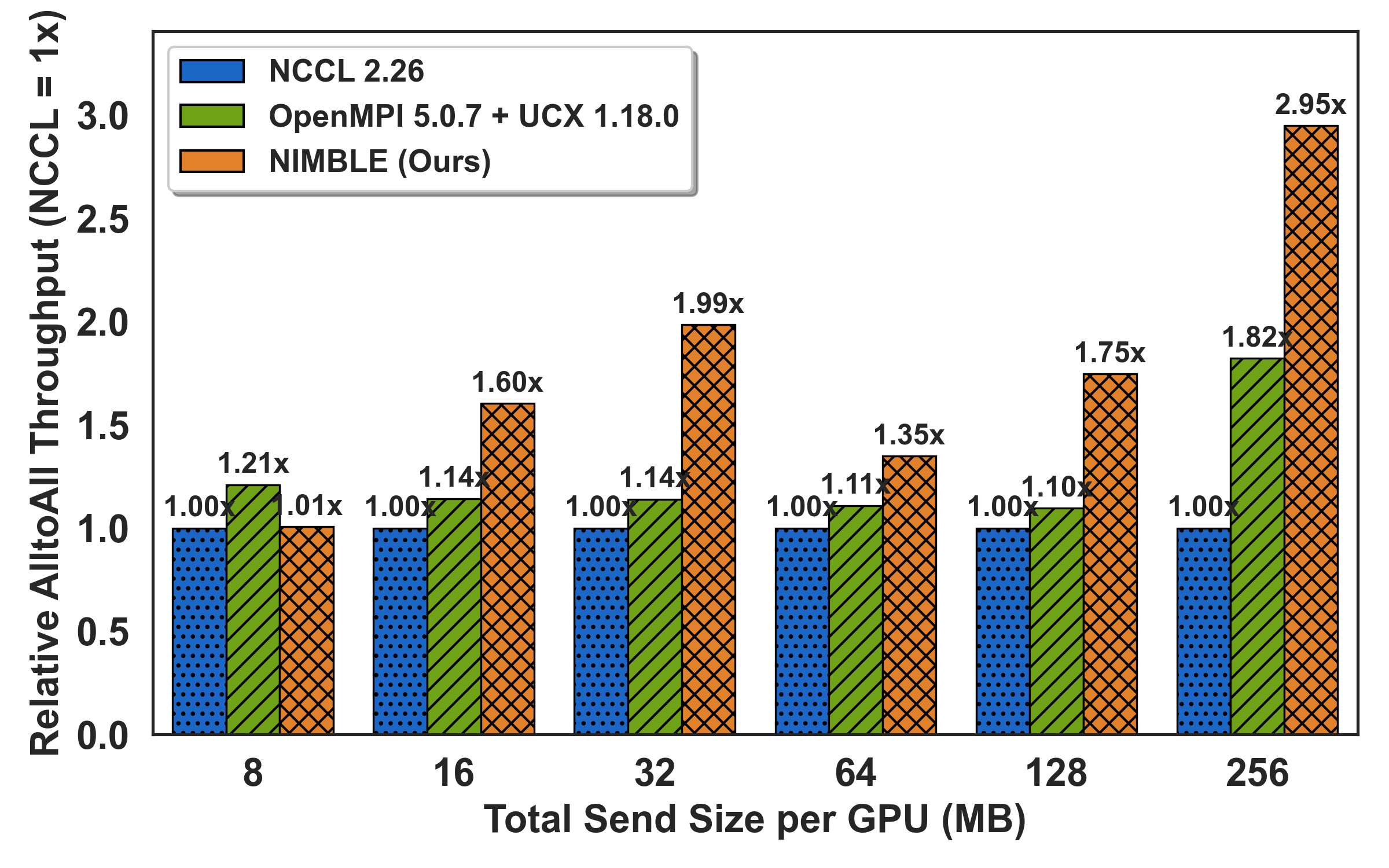}
        \caption{Hotspot ratio=0.7}
    \end{subfigure}
    \hfill
    \begin{subfigure}[t]{0.24\textwidth}
        \includegraphics[width=\linewidth]{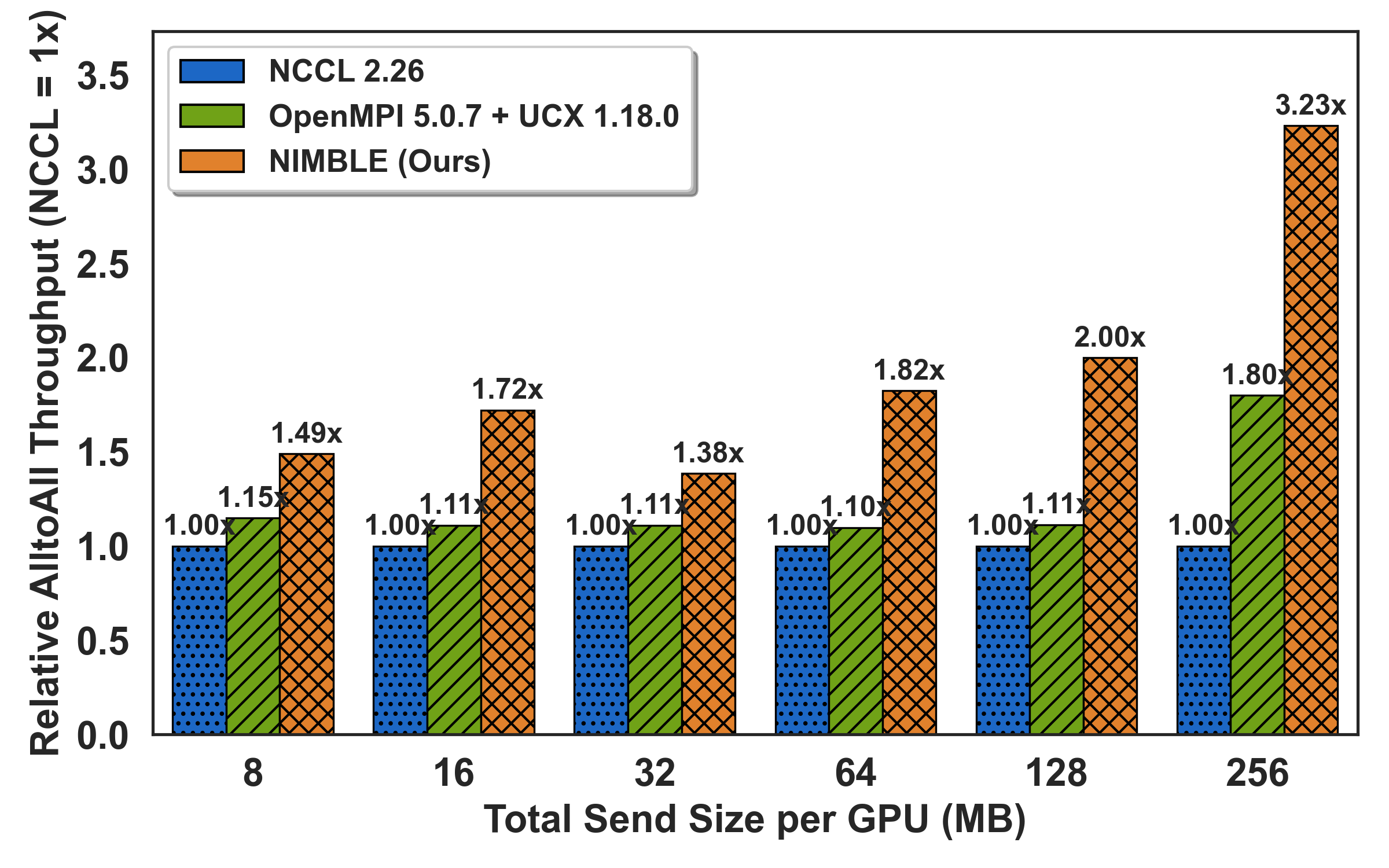}
        \caption{Hotspot ratio=0.8}
    \end{subfigure}%
    \hfill
    \begin{subfigure}[t]{0.24\textwidth}
        \includegraphics[width=\linewidth]{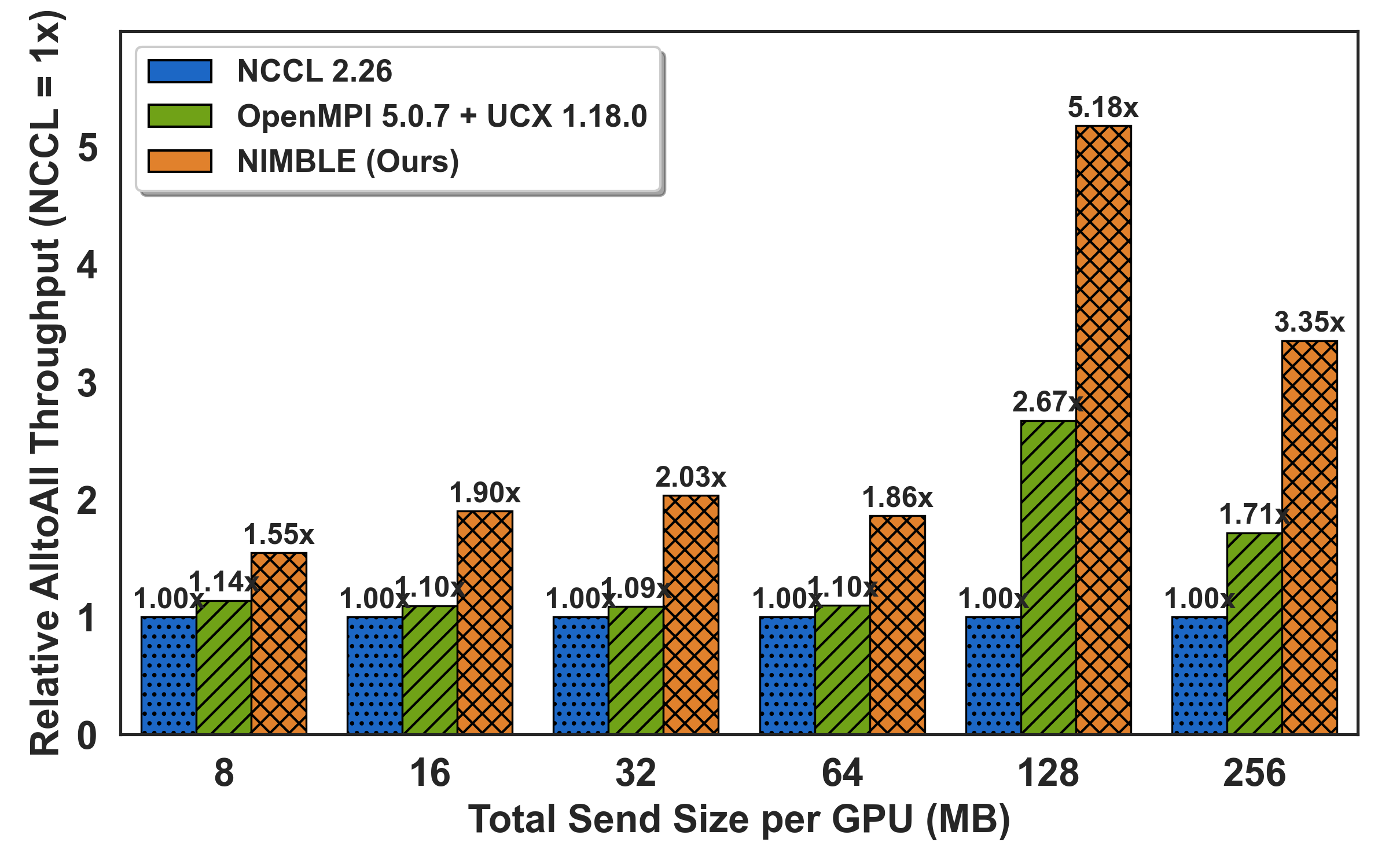}
        \caption{Hotspot ratio=0.9}
    \end{subfigure}%
    \hfill
    \begin{subfigure}[t]{0.24\textwidth}
        \includegraphics[width=\linewidth]{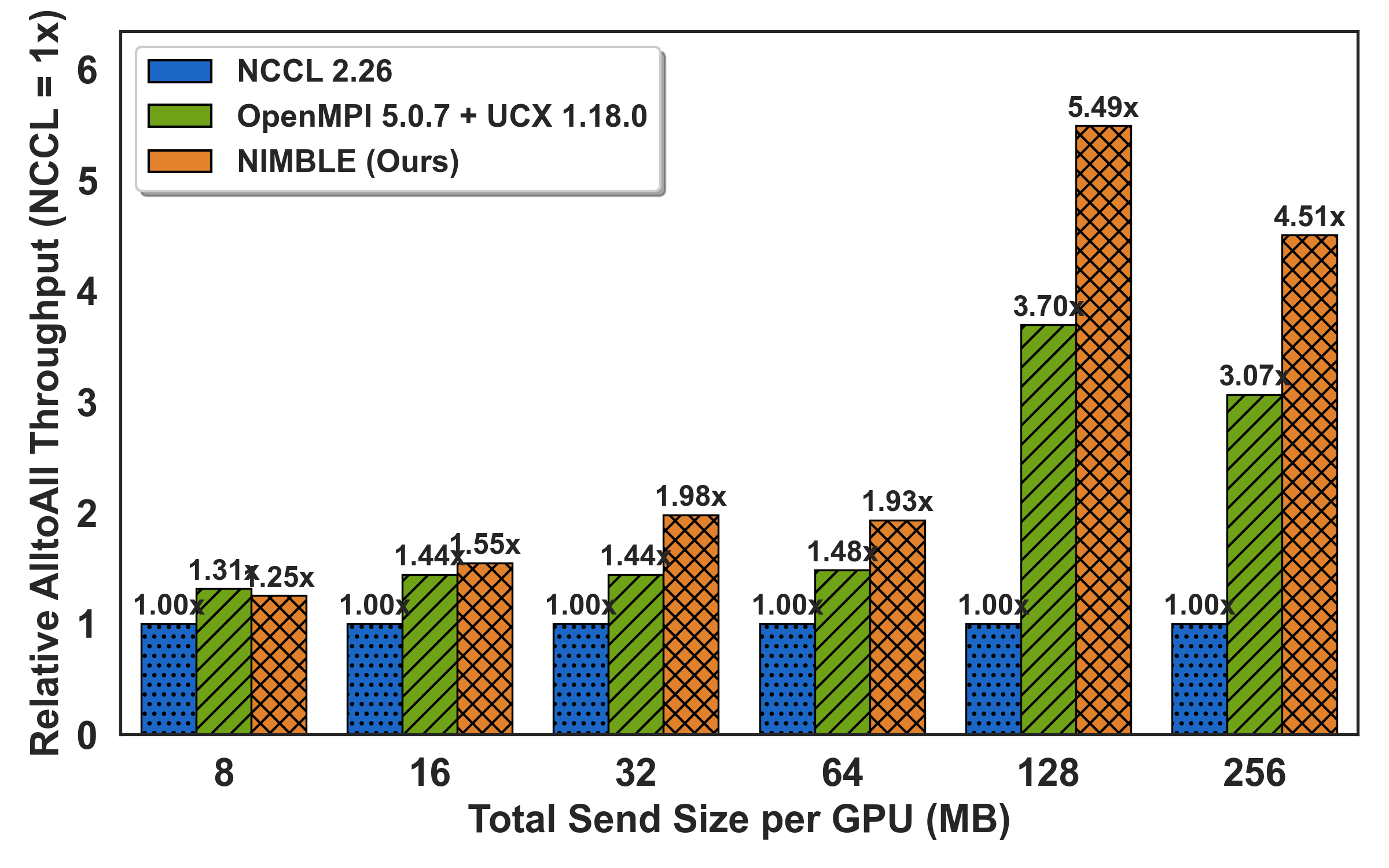}
        \caption{Hotspot ratio=1.0}
    \end{subfigure}

    \caption{Skewed All-to-Allv communication. Each rank sends variable-sized messages to all peers; we model skew by directing a fraction (the hotspot ratio) of each rank’s payload to a single “hot” destination, with the remainder distributed evenly.}
    \label{fig:skew_bw}
    \vspace{-1em}
\end{figure*}

\begin{figure*}[h]
    \centering
    \includegraphics[width=0.92\textwidth]{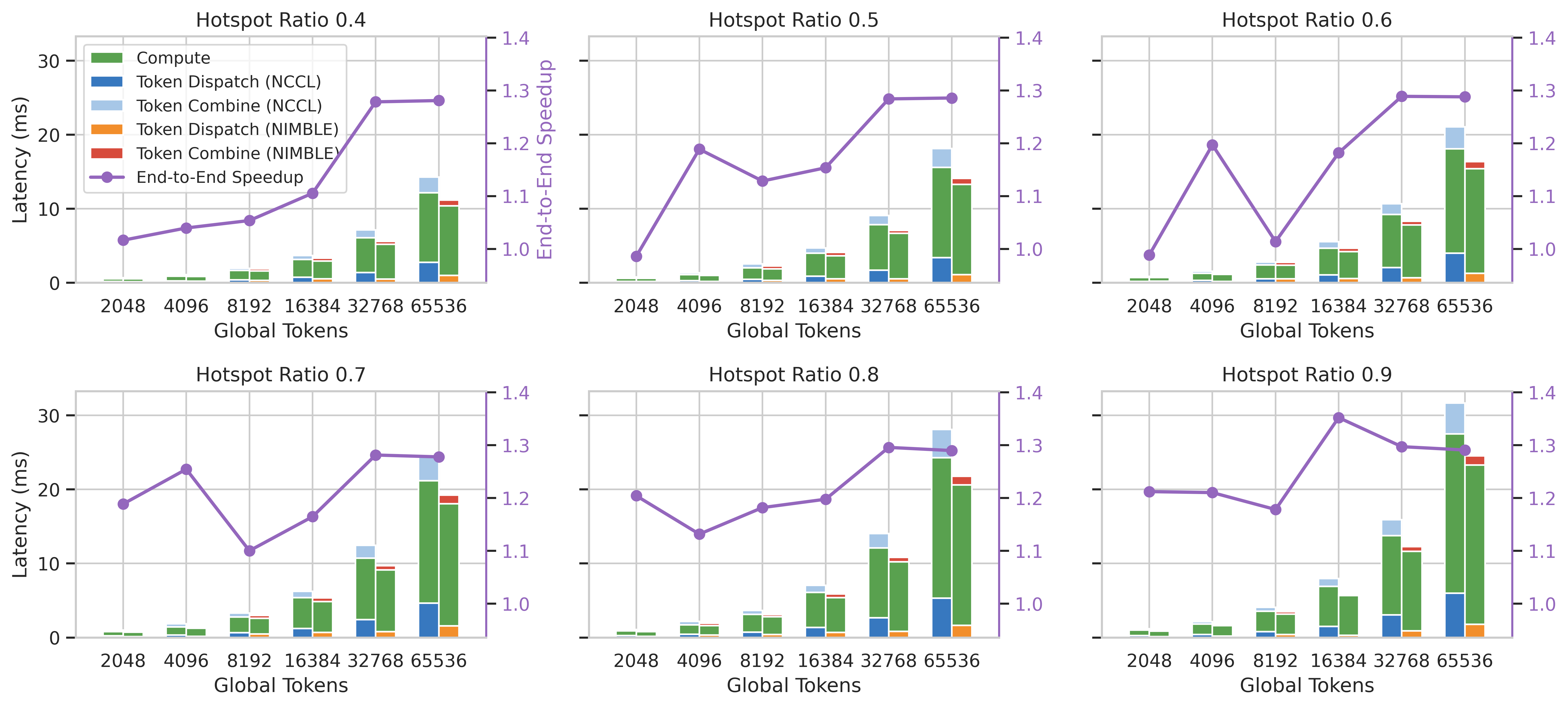}
    \caption{MoE end-to-end latency breakdown for NCCL vs. NIMBLE. For each token size, paired stacks show NCCL (left: blue dispatch, green compute, light-blue combine) and NIMBLE (right: orange dispatch, green compute, red combine); the purple trace overlays end-to-end speedup. Compute is identical across methods; gains come from slimmer dispatch/combine.}
    \label{fig:ep-latency-stack}
\end{figure*}

\subsection{Skewed All-to-Allv}
The All-to-Allv primitive is ubiquitous in large-scale GPU workloads where each rank exchanges variable-sized messages with all peers. In practice, message distributions are often highly skewed: a nontrivial fraction of traffic from many sources concentrates on a small subset of destinations, producing severe per-link imbalance. When such exchanges are launched concurrently across multiple communication groups, the same skew pattern repeats and compounds pressure on NICs and switches. These characteristics motivate adaptive routing and load-aware orchestration.

Figure~\ref{fig:skew_bw} compares overall communication speedup under controlled skew. In this setup with 8 GPUs across two nodes, each GPU directs a fixed fraction of its payload to a designated “hot” peer, while the remaining payload is spread across the other peers. By varying this fraction (the \emph{hotspot ratio}), we evaluate NCCL, OpenMPI, and our NIMBLE under different imbalance levels. When skew is modest and messages are not large, NIMBLE performs on par with NCCL (both kernel-based), while OpenMPI can be slightly better. This is because many MPI stacks drive transfers via GPU DMA engines; such copy-engine–based paths can more easily saturate fabrics at small message sizes than kernel-driven schemes. Accordingly, for small messages NIMBLE avoids multi-path splitting for bandwidth efficiency, as also illustrated in Figure~\ref{fig:four_figures}.

As skew intensifies, NIMBLE progressively outperforms by rerouting traffic off heavily loaded links to less busy paths, substantially reducing tail latency and improving throughput. For example, when the hotspot ratio reaches 0.7 or higher, NIMBLE is up to 5.2$\times$ faster than NCCL.

\subsection{Mixture-of-Experts with expert parallelism}
Deployed mixture-of-experts LLMs commonly use top-$k$ gating at inference without strict capacity caps to avoid token dropping and preserve quality (DeepSeek-V2/V3~\cite{liu2024deepseekv2,liu2024deepseekv3}; Qwen2-MoE~\cite{yang2024qwen2}). DeepSeek-V3~\cite{liu2024deepseekv3} explicitly reports \emph{no token dropping} in both training and inference and instead applies an auxiliary-loss-free, batch-wise balancing bias during training with deployment-time heuristics. Despite such training-time biasing, inference traffic often drifts from training and concentrates on “hot” experts/devices; Huang et~al.~\cite{huang2024toward} identify the gating mechanism as a principal source of inference inefficiency and motivate dynamic gating with expert load balancing. Contemporary systems—Lazarus~\cite{wu2024lazarus}, Pro-Prophet~\cite{wang2024proprophet} address skew via expert replication/placement and cross-device assignment; a late-2024 survey~\cite{liu2024survey-moe-infer} synthesizes these directions and frames inference-time load balancing as a central systems challenge.

We evaluate a two-node, eight-GPU EP configuration (4~GPUs per node) and sweep global token counts \{2\,K, 4\,K, 8\,K, 16\,K, 32\,K, 64\,K\}. Each token has dimension 4096 in \texttt{bfloat16}, and expert compute is a two-layer FFN with 4$\times$ expansion. For each configuration we break down step latency into three phases: \emph{dispatch} (All-to-Allv token routing), \emph{compute} (FFN on received tokens), and \emph{combine} (All-to-Allv gather back to owners). Figure~\ref{fig:ep-latency-stack} presents a stacked per-token latency view: for each token size, the left stack is NCCL (blue dispatch, green compute, light-blue combine), and the right stack is NIMBLE (orange dispatch, green compute, red combine). A purple trace overlays the end-to-end speedup. Compute (green) is identical between methods; thus, all end-to-end gains come from slimmer \emph{dispatch} and \emph{combine} blocks under NIMBLE. As global token count and hotspot ratio increase, dispatch/combine dominate more of the step time, and NIMBLE, correspondingly, yields larger reductions.

Across the sweep, the speedup curve rises almost linearly with token count within each hotspot subplot and shifts upward with higher hotspot ratios: the average speedup climbs from $\sim$1.13$\times$ at hotspot~0.4 to $\sim$1.26$\times$ at hotspot~0.9, capped by a 1.35$\times$ peak at 16\,K tokens when the hotspot consumes 90\% of tokens. The measurements suggest a simple rule: enable NIMBLE once (i) global tokens $\ge$16\,K and (ii) hotspot ratio $\ge$0.7; in this region NIMBLE is consistently $>$1.16$\times$ faster, often $\sim$1.3$\times$. For small, mildly skewed jobs (e.g., 2\,K tokens at hotspot 0.5), prefer the baseline or batch multiple steps before switching to NIMBLE to amortize orchestration overhead.

\subsection{Multi-tenant interference}
\label{sec:multitenant}

In production GPU clusters, large synchronous DL jobs are typically granted \emph{exclusive GPUs/nodes} for predictability and gang-scheduling locality; schedulers emphasize consolidated placement and goodput rather than colocating independent trainings on the same device \cite{xiao2018gandiva,gu2019tiresias,qiao2021pollux}. While GPU sharing exists, it is generally used to harvest slack or pack small/best-effort jobs because interference and preemption overheads can erode tail latency and utilization \cite{xiao2020antman,yu2020salus,strati2024orion}. Consequently, “multiple parallel large trainings on the same node” is atypical; concurrency is primarily across jobs on the shared fabric.

NIMBLE is not a cross-job scheduler; it complements the fabric’s congestion-control layer (e.g., DCQCN, HPCC) that governs fairness and queueing across tenants \cite{zhu2015dcqcn,li2019hpcc}. By re-slicing a job’s traffic over intra-node NVLink edges and inter-node rails based on live link costs, NIMBLE reduces per-job hotspotting and trims communication tails even under background load, while the network’s CC preserves inter-tenant fairness \cite{xiao2018gandiva,gu2019tiresias,qiao2021pollux,xiao2020antman,yu2020salus,strati2024orion,zhu2015dcqcn,li2019hpcc}.

%% file: content/6-Related_works.tex
\section{Related Works}
\label{sec:relat}

For system-level topology-aware communication optimizations, TACCL~\cite{shah2023taccl} introduces a framework for synthesizing efficient collective communication algorithms tailored to specific hardware topologies. By utilizing "communication sketches"—high-level abstractions provided by algorithm designers—TACCL reduces the search space for algorithm synthesis, enabling the generation of optimized communication strategies. The system formulates the synthesis problem as a mixed-integer linear programming (MILP) task, decomposed into routing and scheduling stages, allowing it to scale to multi-node topologies. However, this method requires offline profiling and MILP solving, which cannot adapt to dynamic and skewed workloads. 

Ghazimirsaeed et al.~\cite{ghazimirsaeed2019efficient} propose a collaborative communication mechanism for MPI neighborhood collectives that leverages common neighborhoods among groups of processes to reduce communication stages through message combining. By modeling the communication pattern as a maximum weighted matching problem in distributed hypergraphs, they develop a distributed algorithm to optimize message routing. 

While these approaches effectively optimize neighborhood collectives by exploiting common communication patterns, it primarily focuses on predefined neighborhood structures and collective operations. In -, our proposed system, NIMBLE, addresses the challenges of skewed point-to-point communication patterns in heterogeneous HPC architectures. NIMBLE operates transparently at runtime, dynamically redistributing data across all available links without requiring prior knowledge of the communication pattern or manual intervention. This makes NIMBLE particularly effective for workloads with irregular and asymmetric communication patterns, where traditional collective optimization techniques may fall short.

%% file: content/7-Limitations_FutureWork.tex
\section{Limitations}
\label{sec:limitations}

Despite NIMBLE being superior with the multi-GPU multi-NIC configurations in most HPC clusters, we notice that for NVIDIA DGX machines, where a single node has 8 GPUs with 1 NVSwitch, the intra-node multi-path forward becomes tricky. In such systems, GPUs are not connected in an all-to-all manner with direct links between each pair; instead, all GPUs are connected to a central NVSwitch, where all communication needs to be routed by it. Given that any GPU in such systems has only one NVLink path to NVSwitch, intra-node forward cannot be done as the only available link is already taken by the direct path. However, inter-node multi-NIC path is still feasible as we can always identify load imbalance on these NICs and reroute messages.

%% file: content/8-Conclusion.tex
\section{Conclusion}
\label{sec:conc}

We presented NIMBLE, a runtime system tackling communication skew by dynamically rebalancing traffic across all available hardware links in heterogeneous HPC environments. Operating transparently and adapting in real-time, NIMBLE significantly boosts communication throughput for skewed workloads. This work demonstrates the benefit of link-aware communication orchestration at the system level. As HPC systems grow in heterogeneity, approaches like NIMBLE are crucial for maximizing resource utilization. Future directions include refining the orchestration engine and investigating an InfiniBand GPUDirect Async (IBGDA) based implementation to further reduce overhead and improve latency by minimizing host interaction.